\documentclass[11pt,a4paper]{article}
\usepackage{cite}
\usepackage{graphicx}
\usepackage{amssymb}
\usepackage{amsmath}
\usepackage{amsfonts}
\usepackage{dsfont}
\usepackage{mathtools}
\usepackage{array}
\usepackage{rotating}
\usepackage{bbold,amsfonts}

\usepackage[utf8]{inputenc}
\usepackage{bm}
\usepackage{xcolor}
\usepackage{float}
\usepackage{braket}

\usepackage[height=8.9in,width=6.45in]{geometry}
\usepackage[font=small,labelfont=bf]{caption}
\usepackage[hidelinks]{hyperref}
\usepackage{booktabs,float,slashed}
\usepackage{standalone}
\usepackage{caption}
\usepackage{subcaption}

\numberwithin{equation}{section}

\newcommand{\beq}{\begin{equation}}
\newcommand{\eeq}{\end{equation}}

\newcommand{\overbar}[1]{\mkern 1.5mu\overline{\mkern-1.5mu#1\mkern-1.5mu}\mkern 1.5mu}

\newcommand{\p}{\pi}

\DeclareMathOperator{\Tr}{Tr}
\DeclareMathOperator{\tr}{tr}

\newcommand{\ii}{\mathrm{i}}

\makeatletter
\newcommand*{\letterdef@}{}
\newcommand*{\letterdef}[3]{%
	\def\letterdef@##1{\expandafter\newcommand\csname #1\endcsname{#2{##1}}}%
	\@tfor\@tempa :=#3\do{\expandafter\letterdef@\expandafter{\@tempa}}}
\makeatother
\letterdef{c#1} {\mathcal}{ABCDEFGHIJKLMNOPQRSTUVWXYZ} 
\letterdef{rm#1}{\mathrm} {dDeimM} 

\newcommand{\E}{{\scriptscriptstyle{\mathbf{E}}}}
\newcommand{\D}{{\scriptscriptstyle{\mathbf{D}}}}
\newcommand{\LO}{{\scriptscriptstyle{(\mathrm{LO})}}}
\newcommand{\NLO}{{\scriptscriptstyle{(\mathrm{NLO})}}}

\newdimen\tableauside\tableauside=1.0ex
\newdimen\tableaurule\tableaurule=0.4pt
\newdimen\tableaustep
\def\phantomhrule#1{\hbox{\vbox to0pt{\hrule height\tableaurule
			width#1\vss}}}
\def\phantomvrule#1{\vbox{\hbox to0pt{\vrule width\tableaurule
			height#1\hss}}}
\def\sqr{\vbox{%
		\phantomhrule\tableaustep
		\hbox{\phantomvrule\tableaustep\kern\tableaustep\phantomvrule\tableaustep}%
		\hbox{\vbox{\phantomhrule\tableauside}\kern-\tableaurule}}}
\def\squares#1{\hbox{\count0=#1\noindent\loop\sqr
		\advance\count0 by-1 \ifnum\count0>0\repeat}}
\def\tableau#1{\vcenter{\offinterlineskip
		\tableaustep=\tableauside\advance\tableaustep by-\tableaurule
		\kern\normallineskip\hbox
		{\kern\normallineskip\vbox
			{\gettableau#1 0 }%
			\kern\normallineskip\kern\tableaurule}%
		\kern\normallineskip\kern\tableaurule}}
\def\gettableau#1 {\ifnum#1=0\let\next=\null\else
	\squares{#1}\let\next=\gettableau\fi\next}
\allowdisplaybreaks

\begin{document}
\begin{titlepage}

\begin{flushright}
\small
\texttt{HU-EP-24/21}
\end{flushright}

\vspace*{10mm}
\begin{center}
{\LARGE \bf 
Integrated correlators in a $\mathcal{N}=2$ SYM theory with 
\\[2mm]
fundamental flavors: a matrix-model perspective
}

\vspace*{15mm}

{\Large M. Bill\`o${}^{\,a,b}$, M. Frau${}^{\,a,b}$, A. Lerda${}^{\,c,b}$, A. Pini${}^{\,d}$, P. Vallarino${}^{a,b}$}

\vspace*{8mm}
	
${}^a$ Universit\`a di Torino, Dipartimento di Fisica,\\
			Via P. Giuria 1, I-10125 Torino, Italy
			\vskip 0.3cm
			
${}^b$   I.N.F.N. - sezione di Torino,\\
			Via P. Giuria 1, I-10125 Torino, Italy 
			\vskip 0.3cm
   
${}^c$  Universit\`a del Piemonte Orientale,\\
			Dipartimento di Scienze e Innovazione Tecnologica\\
			Viale T. Michel 11, I-15121 Alessandria, Italy
			\vskip 0.3cm
   
${}^d$ Institut f{\"u}r Physik, Humboldt-Universit{\"a}t zu Berlin,\\
IRIS Geb{\"a}ude, Zum Großen Windkanal 2, 12489 Berlin, Germany  

\vskip 0.8cm
	{\small
		E-mail:
		\texttt{billo,frau,lerda,vallarin@to.infn.it;alessandro.pini@physik.hu-berlin.de}
	}
\vspace*{0.8cm}
\end{center}

\begin{abstract}
The \textbf{D} theory is a $\cN=2$ conformal SYM theory in four dimensions with gauge group $\mathrm{SU}(N)$, four matter hypermultiplets in the fundamental and two in the anti-symmetric representation, and a flavor symmetry that contains a $\mathrm{U}(4)$ factor. Its holographic dual is obtained via a combination of orbifold and orientifold projections from Type II B string theory on $\mathrm{AdS}_5 \times S^5$ and possesses a sector of open strings attached to D7 branes with $\mathrm{U}(4)$ Chan-Paton factors. The $\mathrm{AdS}$ Veneziano amplitude of the $\mathrm{U}(4)$ gluons is dual to four-point correlators of moment-map operators of the $\mathrm{U}(4)$ flavor symmetry in the \textbf{D} theory. An integrated version of these correlators is captured by a deformation of the \textbf{D} theory in which the fundamental hypermultiplets acquire a mass. The partition function of this massive theory can be evaluated with matrix-model techniques using localization. In this paper we analyze the matrix model of the mass-deformed \textbf{D} theory arising from localization using the so-called ``full Lie algebra'' approach in a $1/N$ expansion. In particular, we study the partition function and its mass derivatives up to $O(1/N^2)$ corrections obtaining exact expressions that are valid for all values of the 't Hooft coupling. We also analyze their behavior at strong coupling where our results produce useful constraints on the dual open string scattering amplitudes in AdS.         
\end{abstract}
\vskip 0.5cm
	{
		Keywords: {matrix model, $\mathcal{N}=2$ SYM theory, localization, strong coupling}
	}
\end{titlepage}
\setcounter{tocdepth}{2}
\tableofcontents

\section{Introduction and summary of results}
\label{sec:intro}
Since the first paper on the subject \cite{Binder:2019jwn}, the study of integrated correlators of primary operators in four-dimensional superconformal gauge theories has gained an increasing attention as an ideal playground to explore non-perturbative physics. On the one hand, integrated correlators can be computed in an efficient way by exploiting supersymmetric localization and matrix-model techniques \cite{Pestun:2007rz}\,%
\footnote{See also the collection \cite{Pestun:2016zxk} and references therein.} which, when the number of colors in the gauge theory is large, often lead to exact results valid in all regimes. On the other hand, they can be studied with standard conformal field theory methods and, through the AdS/CFT correspondence, can be used to put constraints on the scattering amplitudes of string modes propagating in AdS \cite{Binder:2019jwn}. 
In the last few years, many aspects of these integrated correlators have been explored in the maximally supersymmetric context of $\cN=4$ super Yang-Mills (SYM) theory \cite{Chester:2019jas,Chester:2020dja,Dorigoni:2021bvj,Dorigoni:2021guq} where modular
\cite{Chester:2020vyz,Collier:2022emf,Dorigoni:2022cua,Paul:2022piq,Dorigoni:2024dhy}, weak-coupling \cite{Wen:2022oky} and non-planar \cite{Zhang:2024ypu} properties have been thoroughly investigated. Further features of integrated correlators have been explored by introducing general gauge groups \cite{Dorigoni:2022zcr,Dorigoni:2023ezg}, by considering operator insertions with generic \cite{Brown:2023zbr} or large conformal dimensions \cite{Paul:2023rka,Brown:2023cpz,Brown:2023why,Caetano:2023zwe,Brown:2024tru}, or by inserting a Wilson line \cite{Pufu:2023vwo,Billo:2023ncz,Billo:2024kri,Dempsey:2024vkf}.

More recently, integrated correlators have also been studied in theories with $\cN=2$ supersymmetry \cite{Chester:2022sqb,Fiol:2023cml,Behan:2023fqq,Billo:2023kak,Alday:2024yax,Behan:2024vwg,Pini:2024uia,Alday:2024ksp}. In particular, in \cite{Behan:2023fqq,Alday:2024yax,Behan:2024vwg,Alday:2024ksp} integrated correlators were used to gain insight on the scattering amplitudes of gluons in an orientifold of Type II B string theory that describes the near-horizon geometry of $N$ D3-branes probing a $D_4$-singularity in F-theory \cite{Sen:1996vd,Banks:1996nj}. Through the AdS/CFT correspondence, this string theory is dual to a four-dimensional $\cN=2$ SYM theory with gauge group\,%
\footnote{In this paper we use the notation Sp($N$) to indicate the symplectic group of rank $N$.} Sp($N$), one anti-symmetric hypermultiplet and four fundamental hypermultiplets with an SO(8) flavor symmetry \cite{Aharony:1998xz}. In the large-$N$ limit, this superconformal theory can be studied efficiently using localization and matrix-model techniques \cite{Beccaria:2021ism,Beccaria:2022kxy} which allow the free energy to be calculated also when the hypermultiplets become massive. The mass-derivatives of the free energy turn out to be particularly interesting quantities because, when the gauge theory is defined on a 4-sphere (as required by the localization procedure) the hypermultiplet mass couples linearly to the so-called moment-map operators. These are the top components of the flavor current multiplet of the $\cN=2$ superconformal algebra \cite{Dolan:2001tt} and, according to the holographic dictionary, are dual to the SO(8) gluons\,%
\footnote{The flavor symmetry of the 4-dimensional conformal field theory becomes a gauge symmetry in the gravitational dual.} of string theory on AdS. Therefore, by computing the fourth mass-derivative of the free energy of the Sp($N$) gauge theory one obtains an integrated 4-point function of moment-map operators which, in the large-$N$ limit, provides information on the dual 4-gluon amplitude in AdS \cite{Behan:2023fqq,Alday:2024yax}. In fact, the form of the latter is constrained by analytic bootstrap 
\cite{Alday:2021odx,Alday:2024yax,Alday:2024ksp} only up to numerical coefficients which are actually fixed by comparing with the localization results.

In this paper, we study along these lines the matrix-model of a different theory, namely the $\cN=2$ SYM theory with gauge group SU($N$), two hypermultiplets in the anti-symmetric representation and four hypermultiplets in the fundamental representation, called \textbf{D} theory \cite{Billo:2019fbi}. With this matter content, the $\beta$-function vanishes and superconformal invariance is present also at the quantum level. In Type II B string theory this model can be engineered with $N$ fractional D3-branes in a $\mathbb{Z}_2$-orbifold probing an O7-orientifold background \cite{Park:1998zh,Ennes:2000fu}. To cancel the Ramond-Ramond charge of the O7-plane and have a consistent configuration, one has to introduce four D7-branes plus their four orientifold images. The massless excitations of the open strings stretching between the D3-branes and the D7-branes give rise to four hypermultiplets in the fundamental representation of SU($N$), while the massless excitations of the open strings on the D3 branes passing through the orientifold plane produce two hypermultiplets in the anti-symmetric of SU($N$). The four-dimensional world-volume theory of the D3-branes has a global symmetry group given by
\begin{align}
    \mathrm{SU}(2)_R\times \mathrm{U}(1)_R \times \mathrm{SU}(2)_L\times \mathrm{U}(1)_L\times \mathrm{U}(4)
    \label{global}
\end{align}
where the first two factors correspond to the $R$-symmetry group of the $\cN=2$ superconformal theory while the second two factors correspond to the flavor symmetry of the two anti-symmetric hypermultiplets. The last factor, U(4), is the flavor symmetry of the four fundamental hypermultiplets and will play an important role in the sequel. From the point of view of the D7-branes, it represents the gauge group of their world-volume theory and arises by taking a $\mathbb{Z}_2$-orbifold projection of the initial SO(8) gauge group of the eight D7-branes in the orientifold background. More precisely, let $\Lambda$ be a hermitian anti-symmetric $8 \times 8$ Chan-Paton matrix in the $\mathfrak{so}(8)$ algebra. Under the $\mathbb{Z}_2$-orbifold it transforms as
\cite{Gimon:1996rq}
\begin{align}
    \Lambda\to \gamma\,\Lambda \,\gamma^{-1} \quad\mbox{with}\quad
    \gamma=\begin{pmatrix}
        \mathbf{0}&-\ii\,\mathbf{1}\\
        \ii\,\mathbf{1}&\mathbf{0}
    \end{pmatrix}~,
    \label{MZ2}
\end{align}
where in bold-face we denote $4\times 4$ blocks. Thus, $\Lambda$ is invariant under the orbifold only if it takes the form
\begin{align}
    \begin{pmatrix}
        \mathbf{A}& \ii\,\mathbf{S}\\
        -\ii\,\mathbf{S}&\mathbf{A}
    \end{pmatrix}\quad \mbox{with}~~~\mathbf{A}^{\!t}=-\mathbf{A}~,~~\mathbf{A}^{\!*}=-\mathbf{A}~,~~\mathbf{S}^t=\mathbf{S}~,~~\mathbf{S}^*=\mathbf{S}~.
    \label{U4}
\end{align}
Matrices of this form represent the embedding into $\mathfrak{so}(8)$ of a $\mathfrak{u}(4)$ hermitian matrix $\mathbf{A} + \mathbf{S}$, so that the Chan-Paton factors (\ref{U4}) actually describe a U(4) group.

This explicit open string construction is instrumental also to discuss the gravitational dual of the \textbf{D} theory which, as discussed for example in \cite{Ennes:2000fu}, is Type II B string theory on $\mathrm{AdS}_5\times S^5/\Gamma$ where $\Gamma$ is a discrete group obtained by combining the orbifold and orientifold $\mathbb{Z}_2$-projections.
The gravitational dual theory contains three sectors: a bulk sector consisting of closed string states that are invariant under the orbifold and orientifold parities and propagate in $\mathrm{AdS}_5\times S^5/\Gamma$, a twisted sector comprising closed string twisted states invariant under the orientifold parity and localized at the $\mathbb{Z}_2$-orbifold fixed locus $\mathrm{AdS}_5\times S^1$, and finally a ``D7-sector'' consisting of open string states invariant under the orbifold and propagating on the world-volume of the eight D7-branes with the topology of $\mathrm{AdS}_5\times S^3$.
Among the states in this last sector, there are the gluons of U(4). According to the holographic dictionary, they are dual to operators of conformal dimension 2 made up with the scalars, $q$ and $\widetilde{q}$, of the fundamental hypermultiplets of the \textbf{D} theory, namely
\begin{align}
    \widetilde{q}\,\boldsymbol{\lambda}^A q \qquad\mbox{and}\qquad q^\dagger\boldsymbol{\lambda}^A\,\widetilde{q}^{\,\dagger}~.
    \label{qtildeq}
\end{align}
Here $\boldsymbol{\lambda}^A$ ($A=1,\ldots,16$) are the generators of U(4) in the fundamental representation, and a sum over the SU($N$) indices has been understood. The two combinations (\ref{qtildeq}) are moment-map operators belonging to the $\cN=2$ flavor current multiplet in the adjoint representation of U(4). This multiplet contains also two scalars, $\Sigma^A$ and $\overbar{\Sigma}^A$, of dimension 3 that are quadratic in the fermionic partners of $q$ and $\widetilde{q}$, and a conserved vector current. 
When the hypermultiplets become massive, the Lagrangian of the \textbf{D} theory on a 4-sphere gets modified by the term\,\footnote{Actually, the massive Lagrangian has also a term proportional to $m_Am^A$ which however does not play any role in what follows \cite{Binder:2019jwn,Chester:2019jas,Chester:2020dja,Dorigoni:2021bvj,Dorigoni:2021guq}. See also the recent discussion of this point in \cite{Dempsey:2024vkf}.}
\begin{align}
   \int d^4x \,\sqrt{g} \,\,m_A\,\Big(\frac{\ii}{r}\,J^A+K^A\Big)
   \label{massivaaction}
\end{align}
where $J^A=\widetilde{q}\,\boldsymbol{\lambda}^A q+q^\dagger\boldsymbol{\lambda}^A\,\widetilde{q}^{\,\dagger}$, $K^A=\Sigma^A+\overbar{\Sigma}^A$, $r$ is the radius of the sphere and $g$ the determinant of its metric.
This massive model will be called \textbf{D}$^*$ theory.

By taking four mass-derivatives of the free energy of the \textbf{D}$^*$ theory and then setting the masses to zero, one finds
\begin{align}
    -\partial_{m_A}\partial_{m_B}\partial_{m_C}\partial_{m_D}
    \cF_{\mathbf{D}^*}\Big|_{m=0}&=\Big\langle
    \int d^4x_1 \,\sqrt{g} \,\Big(\frac{\ii}{r}\,J^A+K^A\Big)\notag\\
    &\qquad \qquad\ldots \int d^4x_4 \,\sqrt{g} \,\Big(\frac{\ii}{r}\,J^D+K^D\Big)\Big\rangle_{\mathbf{D}}~,
    \label{integrated}
\end{align}
namely an integrated correlator of 4 moment-map operators $J$ and their descendants $K$ in the \textbf{D} theory. Since the flavor multiplet is half-BPS, by exploiting superconformal Ward identities \cite{Dolan:2000ut} we can express all correlators in the right-hand side of (\ref{integrated}) only in terms of 4-point functions of the moment-map operators and get
\begin{align}
    -\partial_{m_A}\partial_{m_B}\partial_{m_C}\partial_{m_D}
    \cF_{\mathbf{D}^*}\Big|_{m=0}=
    \int \!\prod_{a=1}^4
    d^4x_a\,\,\mu\big(\{x_a\}\big)\,\big\langle J^A(x_1)\,J^B(x_2)\,J^C(x_3)\,J^D(x_4)\,\big\rangle_{\mathbf{D}}
    \label{integrated2}
\end{align}
where $\mu\big(\{x_a\}\big)$ is a suitable integration measure which has been determined in \cite{Chester:2020dja,Chester:2022sqb}. Through the AdS/CFT correspondence, the 4-point correlators of moment-map operators are dual to the 4-point scattering amplitudes of gluons in AdS, and thus from (\ref{integrated2}) we see that the quartic mass-derivatives of the free-energy yield important information for the integrated scattering amplitudes of gluons in AdS. The left-hand side of (\ref{integrated2}) can be explicitly evaluated using a matrix model provided by supersymmetric localization. This is what we do in this paper working order by order in the $1/N$ expansion.

\subsection{Summary of results}
Although they have similar features from the holographic point of view, the Sp($N$) theory considered in \cite{Behan:2023fqq} and the \textbf{D} theory considered in this paper have different gauge groups, matter contents and flavor groups, and therefore it is highly non-trivial to verify that the two theories have similar strong-coupling behavior. The differences between the two theories are evident also in the associated
matrix models. In fact, the Sp($N$) matrix model is a simple deformation of the free Gaussian model produced by a series of single-trace interactions
(see for example \cite{Beccaria:2021ism,Beccaria:2022kxy}), while the matrix model of the \textbf{D} theory, in addition to single-trace terms, also contains a double-trace part. As we will explicitly show in Section~\ref{sec:Dstar}, this double-trace part coincides with the interaction term of the matrix model associated to another $\cN=2$ superconformal gauge theory, called \textbf{E} theory, whose matter content consists of two hypermultiplets, one in the symmetric and one in the anti-symmetric representation of SU($N$). The \textbf{E} theory and its quiver ancestor have been studied in detail in a series of papers \cite{Beccaria:2020hgy,Galvagno:2020cgq,Beccaria:2021vuc,Beccaria:2021hvt,Billo:2021rdb,Billo:2022xas,Billo:2022gmq,Billo:2022fnb,Beccaria:2022ypy,Billo:2022lrv} where many exact results in the large-$N$ limit have been obtained using the so-called full Lie algebra approach originally introduced in \cite{Billo:2017glv}. Unlike the more standard eigenvalue approach, the full Lie algebra approach, which is based on the use of recursive fusion/fission relations, allows the double-trace interactions to be treated exactly. Relying on this method, we are able to obtain explicit results also for the \textbf{D} theory matrix model. A key ingredient of our analysis is the choice of a particular basis of operators
that leads to a simple expression for the interaction terms of the matrix model which is valid for all values of the coupling constant and which allows one to compute correlation functions in an efficient and systematic way. 

In Section~\ref{D*sec} we turn to the \textbf{D}$^*$ theory by giving a mass to the fundamental hypermultiplets, derive the corresponding matrix model and calculate the quartic mass-derivatives of its free energy. Using the special basis of operators previously introduced, we manage to obtain, order by order in the large-$N$ expansion, an explicit form of these mass-derivatives in terms of integrals of Bessel functions.
In this way we fill the gap of previous investigations which never considered the \textbf{D} theory because of the complicated structure of its matrix model. 

In Section~\ref{secn:strong} we analyze the quartic mass-derivatives at strong coupling where they become very similar (but not identical) to those of the Sp($N$) theory. While perturbatively the \textbf{D} theory and the Sp($N$) theory are completely different, we find that the differences vanish at strong coupling in the large-$N$ limit, implying that the holographic dual of the two theories be of the same type. Finally, in Section\,\ref{sec:concl} we present our conclusions and point out some possible further developments, while in Appendix~\ref{app:strong} we collect some technical details on the strong-coupling analysis and in Appendix~\ref{app:spn} we show that the full Lie algebra approach can be also applied to the matrix model of the Sp($N$) theory, allowing us to recover the results of \cite{Behan:2023fqq} in an alternative, simple way\,%
\footnote{We observe that these techniques can be applied also to study the conformal theory with gauge group Sp($N$)$\times$Sp($N$), one bi-fundamental and four fundamental hypermultiplets, and flavor group SO(4)$\times$SO(4) \cite{Ennes:2000fu}. This model originates from a different orbifold projection of the Sp($N$) theory, and is the only other theory that describes scattering on D7-branes. Its associated matrix model is similar to that of the quiver theories studied using the full Lie-algebra approach in \cite{Billo:2021rdb,Billo:2022gmq,Billo:2022fnb,Billo:2022lrv,Pini:2024uia} and thus it can be analyzed along the same lines.}.

\section{The matrix-model description of \texorpdfstring{$\cN=2$ SYM theories}{}}
\label{sec:Dstar}
Let us consider a $\cN=2$ SU($N$) gauge theory with $N_F$ fundamental, $N_S$ symmetric and $N_A$ anti-symmetric hypermultiplets. If
\begin{align}
    N_F=N(2-N_S-N_A)+2(N_A-N_S)~,
    \label{Nf}
\end{align}
the $\beta$-function vanishes and the theory is superconformal. There are five families of solutions to this requirement for generic $N$ which are displayed in Tab.\,\ref{tab:1}.
\begin{table}[ht]
\begin{center}
\begin{tabular}{ c|c|c|c| }
 & $\phantom{\Big|}N_F$& $\phantom{\Big|}N_S$ & $\phantom{\Big|}N_A$ \\\hline\hline
Theory \textbf{A} (a.k.a. SQCD) & $\phantom{\Big|}2N$& $\phantom{\Big|}0$ & $\phantom{\Big|}0$
\\\hline
Theory \textbf{B}  & $\phantom{\Big|}N-2$& $\phantom{\Big|}1$ & $\phantom{\Big|}0$
\\\hline
Theory \textbf{C}  & $\phantom{\Big|}N+2$& $\phantom{\Big|}0$ & $\phantom{\Big|}1$
\\\hline
Theory \textbf{D}  & $\phantom{\Big|}4$& $\phantom{\Big|}0$ & $\phantom{\Big|}2$
\\\hline
Theory \textbf{E}  & $\phantom{\Big|}0$& $\phantom{\Big|}1$ & $\phantom{\Big|}1$
\\\hline
\end{tabular}
\end{center}
\caption{The five families of $\cN=2$ superconformal theories with group SU($N$).}
\label{tab:1}
\end{table}

Placing any of these theories on a 4-sphere and exploiting supersymmetric localization \cite{Pestun:2007rz} (see also \cite{Pestun:2016zxk} and references therein), we can write the partition function as an integral over a Hermitian matrix $a\in \mathfrak{su}(N)$ according to\,%
\footnote{In the full Lie algebra approach we integrate over all elements of $a$.}
\begin{align}
    \cZ=\Big(\frac{8\pi^2 N}{\lambda}\Big)^{\frac{N^2-1}{2}}\int\!da\,\,\rme^{-\frac{8\pi^2N}{\lambda}\tr a^2 }\,|Z_{\mathrm{1-loop}}\,\,Z_{\mathrm{inst}}|^2
    \label{zmm}
\end{align}
where $\lambda$ is the 't Hooft coupling. In the large-$N$ limit with $\lambda$ fixed, the instanton contributions are exponentially suppressed and one can set $Z_{\mathrm{inst}}=1$. The 1-loop part $Z_{\mathrm{1-loop}}$ depends on the hypermultiplet content. For the theories in Tab.\,\ref{tab:1} it is given by
\begin{align}
\label{Z1loop}
\Big|Z_{\mathrm{1-loop}}\Big|^2 = \frac{ \displaystyle \prod_{u < v}^N H^2(\ii a_u-\ii a_v)}{ \displaystyle 
\bigg[\prod_{i=1}^{N_F}\prod_{u=1}^N H(\ii a_u)\bigg]
\,\bigg[
\prod_{i=1}^{N_S}\prod_{u\leq v}^{N}H(\ii a_u+\ii a_v)\bigg]
\,\bigg[
\prod_{i=1}^{N_A}\prod_{u<v}^{N}H(\ii a_u+\ii a_v)\bigg]}    
\end{align}
where $a_u$ are the eigenvalues of $a$ (subject to the tracelessness condition $\sum_u a_u=0$) and 
\begin{align}
    H(x)=\rme^{(1+\gamma)x^2}G(1+x)G(1-x)
\end{align}
with $\gamma$ being the Euler-Mascheroni constant e $G$ the Barnes $G$-function. Rescaling
\begin{align}
    a\to \sqrt{\frac{\lambda}{8\pi^2N}}\,a
    \label{rescaling}
\end{align}
and then expanding in $\lambda$ using
\begin{align}
\log H(x) = -\sum_{n=1}^{\infty}\frac{\zeta_{2n+1}}{n+1}   \,x^{2n+2}
\end{align}
where $\zeta_n$ is the Riemann $\zeta$-value $\zeta(n)$, we can recast the partition function (\ref{zmm}) in the form
\begin{align}
    \cZ=\int\!da\,\,\rme^{-\tr a^2 }\,\rme^{-S_{\mathrm{int}}}
\end{align}
with \cite{Beccaria:2020hgy}
\begin{align}
    S_{\mathrm{int}}=&\,\frac{N_S+N_A+2}{2}\,
   \sum_{k=1}^{\infty}\sum_{\ell=1}^{k-1}(-1)^{k}
   \Big(\frac{\lambda}{8\pi^2N}\Big)^{k+1}\,\binom{2k+2}{2\ell+1}\,
\frac{\zeta_{2k+1}}{k+1}\,\tr a^{2\ell+1}\, \tr a^{2k-2\ell+1}\notag\\[1mm]
&+\frac{N_S+N_A-2}{2}\,
   \sum_{k=1}^{\infty}\sum_{\ell=1}^{k}(-1)^{k}
   \Big(\frac{\lambda}{8\pi^2N}\Big)^{k+1}\,\binom{2k+2}{2\ell}\,
\frac{\zeta_{2k+1}}{k+1}\,\tr a^{2\ell}\, \tr a^{2k-2\ell+2}\notag\\[1mm]
&+2(N_S-N_A)\sum_{k=1}^{\infty} (-1)^{k} \Big(\frac{\lambda}{8\pi^2N}\Big)^{k+1}\,(2^{2k}-1)\,\frac{\zeta_{2k+1}}{k+1}\,\tr a^{2k+2}~.
\label{Sint}
\end{align}

Of particular interest for us are the \textbf{E} and \textbf{D} theories. For the \textbf{E} theory we have $N_S=N_A=1$, $N_F=0$ and thus the interaction action (\ref{Sint}) reduces to
\begin{align}
    S_\E=2\,
   \sum_{k=1}^{\infty}\sum_{\ell=1}^{k-1}(-1)^{k}
   \Big(\frac{\lambda}{8\pi^2N}\Big)^{k+1}\,\binom{2k+2}{2\ell+1}\,
\frac{\zeta_{2k+1}}{k+1}\,\tr a^{2\ell+1}\, \tr a^{2k-2\ell+1}~.
\label{SintE}
\end{align}
This is a sum of double traces of only odd powers of $a$.
For the \textbf{D} theory instead, we have $N_S=0$, $N_A=2$ and $N_F=4$. This implies that the interaction action in this case is
\begin{align}
    S_\D=S_\E+S_{\mathrm{s.t}}
    \label{SintD}
    \end{align}
where $S_{\mathrm{s.t}}$ is the following single-trace term
\begin{align}
    S_{\mathrm{s.t.}}=4\sum_{k=1}^{\infty} (-1)^{k+1} \Big(\frac{\lambda}{8\pi^2N}\Big)^{k+1}\,(2^{2k}-1)\,\frac{\zeta_{2k+1}}{k+1}\,\tr a^{2k+2}~.
\label{Sintst}
\end{align}
Given this structure, we can regard the matrix model of the \textbf{E} theory as a 
``perturbation'' of the free Gaussian model representing $\cN=4$ SYM by means of $S_\E$, and in turn regard the matrix model of the \textbf{D} theory as a ``perturbation'' of the \textbf{E} theory by means of $S_{\mathrm{s.t.}}$. In other words, in the \textbf{E} theory we write the vacuum expectation value of any operator $\cO$ as
\begin{align}
    \big\langle \cO \big\rangle_\E=\frac{\big\langle \cO\,\rme^{-S_\E}\big\rangle_0\phantom{\Big|}}{\big\langle \rme^{-S_\E}\big\rangle_0\phantom{\Big|}}
    \label{vevE}
\end{align}
where $\langle\,\rangle_0$ denotes the vacuum expectation value in the free Gaussian model ({\it{i.e.}} $\cN=4$ SYM), while in the \textbf{D} theory we have
\begin{align}
    \big\langle \cO \big\rangle_\D=\frac{\big\langle \cO\,\rme^{-S_{\mathrm{s.t.}}}\big\rangle_\E\phantom{\Big|}}{\big\langle \rme^{-S_{\mathrm{s.t.}}}\big\rangle_\E\phantom{\Big|}}~.
    \label{vevD}
\end{align}
Thus, ultimately everything is reduced to the computation of correlators in the free Gaussian matrix model. In \cite{Beccaria:2021hvt,Beccaria:2021vuc,Billo:2022xas,Beccaria:2022ypy,Billo:2023kak} it was shown that in the \textbf{E} theory the interaction action $S_\E$ can be treated exactly order by order in the large-$N$ expansion obtaining expressions valid for all values of $\lambda$ written in terms of convolutions of Bessel functions. Here we show that this is possible also in the \textbf{D} theory since at any given order in the large-$N$ expansion only a \emph{finite} number of insertions of $S_{\mathrm{s.t.}}$ are needed. This is the main reason while, despite the complicated structure of the matrix model of the \textbf{D} theory, one can obtain exact results in closed form in the large-$N$ expansion.

Let us now give some details. The first step is to change basis and instead of using $\tr a^{k}$ we introduce a new set of operators $\cP_k$ that are orthonormal in the planar limit of the Gaussian model. They are defined as
\begin{align}
    \cP_k=\sqrt{k}\,\sum_{\ell=0}^{\lfloor \frac{k-1}{2}\rfloor}(-1)^\ell\,\Big(\frac{N}{2}\Big)^{\ell-\frac{k}{2}}\,
    \frac{(k-\ell-1)!}{\ell!\,(k-2\ell)!}\,\Big(\tr a^{k-2\ell}-
    \big\langle \tr a^{k-2\ell}\big\rangle_0\,\Big)~.
    \label{Pkis}
\end{align}
Up to an overall normalization, these operators are related to those proposed in \cite{Rodriguez-Gomez:2016cem} in terms of the Chebyshev polynomials of the first kind $T_k$. Indeed, one can verify that
\begin{align}
   \frac{\sqrt{k}}{2} \,\cP_k=T_k\Big(\frac{x}{\sqrt{2N}}\Big)-T_k(0)
\end{align}
where in the right hand side one has to make the replacement
$x^\ell\to \tr a^\ell-\langle \tr a^\ell\rangle_0$. By inverting (\ref{Pkis}) we get
\begin{align}
\label{changebasis}
\text{tr}\,a^k = \left(\frac{N}{2}\right)^{\frac{k}{2}}\sum_{\ell=0}^{\lfloor\frac{k-1}{2}\rfloor}\sqrt{k-2\ell}\, \binom{k}{\ell}\,\mathcal{P}_{k-2\ell}   + \langle \text{tr}a^k \rangle_0~.
\end{align}
Since $\langle \tr a^{2k+1}\rangle_0=0$, the odd traces are linear combinations of odd $\cP$'s without extra terms; on the contrary the even traces are replaced by a linear combination of even $\cP$'s plus an operator-independent term. Performing this change of basis, we can rewrite the interaction action of the \textbf{E} theory as
\begin{align}
S_\E = -\frac{1}{2}\sum_{k,\ell=1}^\infty\cP_{2k+1} \,\mathsf{X}_{k,\ell} \,\cP_{2\ell+1}
\label{SintE1}
\end{align}
where, as shown in \cite{Beccaria:2020hgy,Beccaria:2021hvt}, the coefficients are given by the following convolution of Bessel functions
\begin{align}
\mathsf{X}_{k,\ell} \,=\, 8\,(-1)^{k+\ell+1}\sqrt{(2k+1)\,(2\ell+1)}\int_0^{\infty}\! \frac{dt}{t}\,\frac{\rme^t}{(\rme^t-1)^2}\,J_{2k+1}\Big(\frac{t\sqrt{\lambda}}{2\pi}\Big)J_{2\ell+1}\Big(\frac{t\sqrt{\lambda}}{2\pi}\Big)~.
\label{Xmatrix}
\end{align}
Notice that even if initially in (\ref{SintE}) $S_\E$ was given as a weak-coupling expansion in powers of $\lambda$, the final expression (\ref{SintE1}) is exact in $\lambda$ and thus it can be used also to study the strong-coupling regime. This form of the interaction action is particularly useful to compute the free energy of the \textbf{E} theory which turns out to be
\begin{align}
    \cF_\E=\frac{1}{2}\Tr \log \big(1-\mathsf{X}\big)+O\Big(\frac{1}{N^2}\Big)~.
    \label{freeenergyE}
\end{align}

Also the single trace term $S_{\mathrm{s.t.}}$ in (\ref{Sintst}) can be conveniently written using the
$\cP$ operators. Inserting (\ref{changebasis}), after simple algebra we get
\begin{align}
    S_{\mathrm{s.t.}} &=  2\sum_{k=1}^{\infty}\sum_{\ell=0}^{k} (-1)^{k+1} \Big(\frac{\lambda}{4\pi^2}\Big)^{k+1}\,(1- 2^{-2k})\,\zeta_{2k+1}
    \frac{\sqrt{2k+2-2\ell}\,(2k+1)!}{\ell!(2k+2-\ell)!}\,\cP_{2k+2-2\ell} \notag\\
    &~+\sum_{k=1}^{\infty} (-1)^{k+1} \Big(\frac{\lambda}{2\pi^2N}\Big)^{k+1}\,(1-2^{-2k})\,\frac{\zeta_{2k+1}}{k+1}\,\big\langle \tr a^{2k+2}\big\rangle_0~.
    \label{Sst1}
\end{align}
The second line is a constant term which does not depend on the matrix operators and contributes only to the overall normalization of the partition function of the \textbf{D} theory. This constant is relevant to compute the free energy $\cF_\D$ but cancels out in the expectation values (\ref{vevD}).
Since our primary interest in this paper is on these expectation values, we will drop this constant term and only consider the first line of (\ref{Sst1}). Exploiting the identity
\begin{align}
\label{zetaidentity}
(1- 2^{-2k}) \,\zeta_{2k+1}= \eta_{2k+1} =\frac{1}{(2k+1)!}
\int_0^{\infty} \frac{dt}{t}\,\frac{\rme^t}{(\rme^t+1)^2}\, t^{2k+2} ~,
\end{align}
where $\eta_n$ is the Dirichlet-$\eta$ value $\eta(n)$, we obtain
\begin{align}
    S_{\mathrm{s.t.}} &=  2\sum_{k=1}^{\infty}\sum_{\ell=0}^{k} (-1)^{k+1} 
    \!\int_0^{\infty}\!\frac{dt}{t}\,\frac{\rme^t}{(\rme^t+1)^2} 
    \Big(\frac{\sqrt{\lambda}\,t}{2\pi}\Big)^{2k+2}
    \frac{\sqrt{2k+2-2\ell}}{\ell!(2k+2-\ell)!}\,\cP_{2k+2-2\ell}~.
    \label{Sst2}
\end{align}
This series can be resummed in terms of Bessel functions. Indeed, after renaming indices, one finds
\begin{align}
\label{Sst3}
S_{\mathrm{s.t.}} = -\sum_{k=1}^{\infty}\mathsf{Y}_{2k}\,\mathcal{P}_{2k}
\end{align}
where
\begin{align}
\label{Yvector}
\mathsf{Y}_{2k} =  (-1)^{k+1} \,2\sqrt{2k}\int_0^\infty \!\frac{dt}{t}\,\frac{\rme^t}{(\rme^t+1)^2}\,J_{2k}\Big(\frac{\sqrt{\lambda}\,t}{\pi}\Big) -\delta_{k,1}\,\frac{\sqrt{2}\log 2}{4\pi^2}\,\lambda~.
\end{align}
In conclusion, the interaction action of the matrix model which is relevant for computing expectation values in the \textbf{D} theory is simply 
\begin{align}
    S_\D= -\frac{1}{2}\sum_{k,\ell=1}^\infty\cP_{2k+1} \,\mathsf{X}_{k,\ell} \,\cP_{2\ell+1}
    -\sum_{k=1}^{\infty}\mathsf{Y}_{2k}\,\mathcal{P}_{2k}
\end{align}
where the $N$-independent coefficients $\mathsf{X}_{k,\ell}$ and $\mathsf{Y}_{2k}$ are explicitly known in terms of integrals of Bessel functions for any value of $\lambda$.

\subsection{The correlators of the \texorpdfstring{$\cP$}{} operators}
Now we have all the ingredients to explicitly compute the correlators of the $\cP$ operators introduced in
(\ref{Pkis}). First we recall a few known facts about such correlators in the free Gaussian matrix model ($\cN=4$ SYM) and in the \textbf{E} theory (see for instance \cite{Billo:2017glv,Billo:2022xas,Billo:2023kak}).

In the large-$N$ expansion, for both $\cN=4$ SYM and the \textbf{E} theory, the $n$-point {\emph{connected}} correlator of the $\mathcal{P}$ operators has the following structure
\begin{align}
    \big\langle \cP_{k_1}\cdots \cP_{k_n}\big\rangle^{(c)} &=
    \frac{f^\LO(k_1,\ldots,k_n)}{N^{n-2}}+\frac{f^\NLO(k_1,\ldots,k_n)}{N^{n}}
    +O\Big(\frac{1}{N^{n+2}}\Big)
    \label{npoint}
\end{align}
where the coefficients $f^\LO$, $f^\NLO$, and the analogous ones at higher sub-leading orders in the large-$N$ expansion, always vanish unless $k_1+\ldots+k_n$ is even. These coefficients are in general functions of the 't Hooft coupling and their explicit expression depends on the theory considered. 
Let us now consider these correlators in more detail starting from $\cN=4$ SYM.

\subsubsection*{\texorpdfstring{$\cN$=4}{} SYM theory}
In $\cN=4$ SYM the 1-point function identically vanishes:
\begin{align}
    \big\langle \cP_k \big\rangle_0=0~.
\end{align}
Thus, for any $k$ we have $f^\LO_{0}(k)=f^\NLO_{0}(k)=\cdots=0$.
For the 2-point correlators, instead, one can show that\,%
\footnote{Here and in the following we always understand that $\sum_i k_i$ must be even.}
\begin{align}
    f^\LO_0(k_1,k_2)=\delta_{k_1,k_2}~.
\end{align}
Indeed, the $\cP$ operators were defined to be orthonormal in the planar limit of the free Gaussian model.
The subleading coefficients can be worked out recursively using the fusion/fission identities of SU($N$) \cite{Billo:2017glv}. For example, at NLO one finds \cite{Pini:2024uia}
    \begin{align}
    f^{\NLO}_0(2k_1,2k_2)&=\frac{\sqrt{2k_1}\sqrt{2k_2}\,(k_1^2+k_2^2-1)(k_1^2+k_2^2-14)}{24}~,\\[1mm]
    f^{\NLO}_0(2k_1+1,2k_2+1)&=\frac{\sqrt{(2k_1+1)}\sqrt{(2k_2+1)}\,(k_1^2+k_2^2+k_1+k_2)(k_1^2+k_2^2+k_1+k_2-14)}{24}~.\notag
\end{align}
However, since in this paper we will work at order $1/N$, we will not use these sub-leading coefficients which indeed appear at order $1/N^2$. Also for the 3-point functions, which scale as $1/N$, the relevant coefficient is the leading one, given by
\begin{align}
    f^{\LO}_0(k_1,k_2,k_3)=\sqrt{k_1}\sqrt{k_2}\sqrt{k_3}~.
\end{align}
In a similar way one can compute the coefficients of the higher point correlators, which however appear at
higher orders in the large-$N$ expansion and thus will not be needed.

\subsubsection*{\textbf{E} theory}
In the \textbf{E} theory, $\cP_{2k}$ and $\cP_{2k+1}$ behave differently since the interaction action of the matrix model only contains
odd operators. In \cite{Billo:2023kak} it was shown that the 1-point function $\langle \cP_{2k}\rangle_\E$ becomes non-trivial at NLO ({\it{i.e.}} at order $1/N$):
\begin{align}
f^\LO_\E(2k)=0~,\qquad
    f^\NLO_\E(2k)=-\sqrt{2k}\,\lambda\,\partial_\lambda \cF_\E
\end{align}
where $\cF_\E$ is the free energy (\ref{freeenergyE}), while $\langle \cP_{2k+1}\rangle_\E$ remains identically zero: $f^\LO_\E(2k+1)=f^\NLO_\E(2k+1)=\cdots=0$.

The 2-point correlators in the \textbf{E} theory were studied in detail in \cite{Beccaria:2021hvt}. Translating these results in the notation of (\ref{npoint}), we have
\begin{equation}
    \begin{aligned}
        f^\LO_\E(2k_1,2k_2)&=\delta_{k_1,k_2}~,\\
        f^\LO_\E(2k_1+1,2k_2+1)&=\mathsf{D}_{k_1,k_2}
    \end{aligned}
\end{equation}
where
\begin{align}
    \mathsf{D}_{k_1,k_2}=\Big(\frac{1}{1-\mathsf{X}}\Big)_{k_1,k_2}
\end{align}
with $\mathsf{X}$ being the matrix defined in (\ref{Xmatrix}). The 3-point correlators have been explicitly computed in \cite{Billo:2022xas} where it was found that
\begin{equation}
    \begin{aligned}
    f^\LO_\E(2k_1,2k_2,2k_3)&=\sqrt{2k_1}\sqrt{2k_2}\sqrt{2k_3}~,\\
    f^\LO_\E(2k_1,2k_2+1,2k_3+1)&=\sqrt{2k_1}\,\mathsf{d}_{k_2}\,\mathsf{d}_{k_3}
\end{aligned}
\end{equation}
where $\mathsf{d}_{k}=\sum_\ell \sqrt{2\ell+1}\,\mathsf{D}_{\ell,k}$.

In Tab.\,\ref{table:Pcorr} we have collected the expressions of the 1-, 2- and 3-point correlators of the even operators which will be used later on. We see that up to the order $1/N$ at which we work, the difference between $\cN=4$ SYM and the \textbf{E} theory is minimal.

\begin{table}[ht]
\begin{center}
\begin{tabular}{ |c||c| }
\hline
$\phantom{\Big|}\cN=4$ SYM & \textbf{E} theory\\\hline\hline
$\phantom{\bigg|}\big\langle \cP_{2k}\big\rangle_0=0$ & $\phantom{\bigg|}\big\langle \cP_{2k}\big\rangle_\E=
-\frac{\sqrt{2k}\,\lambda\,\partial_\lambda \cF_\E}{N}+O\!\left(\frac{1}{N^3}\right)$ \\\hline
$\phantom{\bigg|}\big\langle \cP_{2k_1} \cP_{2k_2}\big\rangle_0=\delta_{k_1,k_2}+O\!\left(\frac{1}{N^2}\right)$ &
$\phantom{\bigg|}\big\langle \cP_{2k_1} \cP_{2k_2}\big\rangle^c_\E=\delta_{k_1,k_2}+O\!\left(\frac{1}{N^2}\right)$\\\hline
$\phantom{\bigg|}\big\langle \cP_{2k_1} \cP_{2k_2} \cP_{2k_3}\big\rangle_0=\frac{\sqrt{2k_1}\sqrt{2k_2}\sqrt{2k_3}}{N}+O\!\left(\frac{1}{N^3}\right)$ &
$\phantom{\bigg|}\big\langle \cP_{2k_1} \cP_{2k_2} \cP_{2k_3}\big\rangle^c_\E=\frac{\sqrt{2k_1}\sqrt{2k_2}\sqrt{2k_3}}{N}+O\!\left(\frac{1}{N^3}\right)$\\\hline
\end{tabular}
\end{center}
\caption{The 1-, 2- and 3-point correlators of even $\cP$'s in the free Gaussian matrix model corresponding to $\cN=4$ SYM and in the matrix model of the \textbf{E} theory up to order $1/N$.}
\label{table:Pcorr}
\end{table}

\subsubsection*{\textbf{D} theory}
Let us now consider the correlators in the \textbf{D} theory starting from the 1-point function of the even operators $\cP_{2n}$. Using the definition (\ref{vevD}) with (\ref{Sst1}), this
1-point function is
\begin{align}
    \big\langle \cP_{2n} \big\rangle_\D =\frac{\Big\langle \cP_{2n}\,\exp\Big({\sum_k \mathsf{Y}_{2k}\,\cP_{2k}}\Big)\Big\rangle_\E\phantom{\bigg|}}{\Big\langle \exp\Big({\sum_k \mathsf{Y}_{2k}}\,\cP_{2k}\Big)\Big\rangle_\E\phantom{\bigg|}} ~.
    \label{P2nD}
\end{align}
Expanding in $\mathsf{Y}_{2k}$, we get
\begin{align}
  \big\langle \cP_{2n} \big\rangle_\D =\big\langle \cP_{2n} \big\rangle_\E+\sum_{k=1}^\infty \mathsf{Y}_{2k} \,\big\langle \cP_{2n}\,\cP_{2k}\rangle_\E^c+ \frac{1}{2}\sum_{k,\ell=1}^\infty \mathsf{Y}_{2k}\,\mathsf{Y}_{2\ell}\,
 \big\langle \cP_{2n}\, \cP_{2k}\, \cP_{2\ell}\rangle_\E^c+\cdots
 \label{P2nD1}
  \end{align}
Exploiting the results for the connected correlators in the \textbf{E} theory that we have recalled above, we see that the first and third terms of (\ref{P2nD1}) are of order $1/N$, while the second term is of order $N^0$. The next terms with more factors of $\mathsf{Y}_{2k}$ are even more sub-leading in the large-$N$ expansion: indeed a term with $q$ such factors is associated to a $(q+1)$-point connected correlator which is of order $1/N^{q-1}$ (for $q\geq2$). This shows that only a {\it{finite}} number of terms in (\ref{P2nD1}) have to be considered up to a given order in the large-$N$ expansion. In particular, as already remarked, the terms explicitly exhibited in $\eqref{P2nD1}$ are the relevant ones to order $1/N$. Thus, we have
\begin{align}
  \big\langle \cP_{2n} \big\rangle_\D =\mathsf{Y}_{2n}+\frac{\sqrt{2n}}{2N}\,\Big({\mathsf{Y}}^{\,2}-2\lambda\,\partial_\lambda\cF_\E\Big)+O\Big(\frac{1}{N^2}\Big)
  \label{P2nD2}
\end{align}
where we have used the \textbf{E} theory connected correlators of Tab.\,\ref{tab:1} and defined\,%
\footnote{The second step in (\ref{Yhat}) directly follows from the recursion relations of the Bessel functions which imply
\begin{align}
    \sum_{\ell=1}^\infty(-1)^{\ell} (2\ell)\,J_{2\ell}(x)=-\frac{x}{2}\,J_1(x)~.
    \label{sumBessel}
\end{align}}
\begin{align}
    {\mathsf{Y}}\,\equiv\,\sum_{k=1}^\infty \sqrt{2k}\,\mathsf{Y}_{2k}=
    \int_0^\infty \frac{dt}{t}\,\frac{e^t}{(e^t+1)^2}\,
\bigg[\frac{\sqrt{\lambda}\,t}{\pi}\,J_{1}\Big(\frac{\sqrt{\lambda}\,t}{\pi}\Big)\bigg]-\frac{\log 2}{2\pi^2}\,\lambda~.
\label{Yhat}
\end{align}
Proceeding in a similar way for the 2-point correlators, we find
\begin{align}
    \big\langle \cP_{2n}\,\cP_{2m} \big\rangle_\D &= \big\langle \cP_{2n}\,\cP_{2m} \big\rangle_\E^c+
    \big\langle \cP_{2n}\big\rangle_\E\,\big\langle \cP_{2m} \big\rangle_\E \notag\\[1mm]
    &~+\sum_{k=1}^\infty \mathsf{Y}_{2k}\bigg[\big\langle \cP_{2n}\,\cP_{2m}\,\cP_{2k}\big\rangle_\E^c+\big\langle \cP_{2n}\big\rangle_\E\,\big\langle \cP_{2m}\,\cP_{2k}\big\rangle_\E^c+\big\langle \cP_{2m}\big\rangle_\E\,\big\langle \cP_{2n}\,\cP_{2k}\big\rangle_\E^c\bigg]\notag\\
    &~+\frac{1}{2}\sum_{k,\ell=1}^\infty \mathsf{Y}_{2k}\,\mathsf{Y}_{2\ell}\bigg[\big\langle \cP_{2n}\,\cP_{2m}\,\cP_{2k}\,\cP_{2\ell}\big\rangle_\E^c+\big\langle \cP_{2n}\big\rangle_\E\,\big\langle \cP_{2m}\,\cP_{2k}\,\cP_{2\ell}\big\rangle_\E^c \notag\\[-1mm]
    &\qquad\qquad\qquad+\big\langle \cP_{2m}\big\rangle_\E\,\big\langle \cP_{2n}\,\cP_{2k}\,\cP_{2\ell}\big\rangle_\E^c
    +2\, \big\langle \cP_{2n}\,\cP_{2k}\big\rangle_\E^c\,\big\langle \cP_{2m}\,\cP_{2\ell}\big\rangle_\E^c\bigg] \notag\\
    &~+\frac{1}{6}\sum_{k,\ell,p=1}^\infty \mathsf{Y}_{2k}\,\mathsf{Y}_{2\ell}\,\mathsf{Y}_{2p}\bigg[\big\langle \cP_{2n}\,\cP_{2m}\,\cP_{2k}\,\cP_{2\ell}\,P_{2p}\big\rangle_\E^c+\big\langle \cP_{2n}\big\rangle_\E\,\big\langle \cP_{2m}\,\cP_{2k}\,\cP_{2\ell}\,\cP_{2p}\big\rangle_\E^c \notag\\
    &\qquad\qquad\qquad+\big\langle \cP_{2m}\big\rangle_\E\,\big\langle \cP_{2n}\,\cP_{2k}\,\cP_{2\ell}\,\cP_{2p}\big\rangle_\E^c+3 \,\big\langle \cP_{2n}\,\cP_{2k}\big\rangle_\E^c\,\big\langle \cP_{2m}\,\cP_{2\ell}\,\cP_{2p}\big\rangle_\E^c\notag\\[2mm]
    &\qquad\qquad\qquad+3 \,\big\langle \cP_{2m}\,\cP_{2k}\big\rangle_\E^c\,\big\langle \cP_{2n}\,\cP_{2\ell}\,\cP_{2p}\big\rangle_\E^c\bigg] + \cdots
    \label{P2nP2m}
\end{align}
Inserting in the right-hand side the expressions of the connected correlators of the \textbf{E} theory and keeping only terms up to order $1/N$, we find
\begin{align}
    \big\langle \cP_{2n}\,\cP_{2m} \big\rangle_\D &=\delta_{n,m}+\mathsf{Y}_{2n}\,\mathsf{Y}_{2m}+\frac{1}{N}\bigg[
    \sqrt{2n}\,\sqrt{2m}\,\,{\mathsf{Y}}
    \notag\\&\qquad\qquad
    +\frac{1}{2}\Big(\sqrt{2n}\,\mathsf{Y}_{2m}+\sqrt{2m}\,\mathsf{Y}_{2n}\Big)\Big({\mathsf{Y}}^{\,2}-2\lambda\,\partial_\lambda\cF_\E\Big)\bigg]+O\Big(\frac{1}{N^2}\Big)~.
    \label{P2nP2m1}
\end{align}
Thus, the connected 2-point function in the \textbf{D} theory is
\begin{align}
    \big\langle \cP_{2n}\,\cP_{2m} \big\rangle_\D^c=\delta_{n,m}+
    \frac{\sqrt{2n}\,\sqrt{2m}\,\,{\mathsf{Y}}}{N}+O\Big(\frac{1}{N^2}\Big)~.
    \label{P2nP2mc}
\end{align}
We stress once again that this expression is exact in $\lambda$.

The 1-point correlator (\ref{P2nD2}) and the connected 2-point correlator (\ref{P2nP2mc}) are the needed ingredients for the purposes of this paper, but of course the techniques we have described can be used to compute higher point correlators, as well as terms of higher order in the large-$N$ expansion.

\section{The matrix model for the \texorpdfstring{$\mathbf{D}^*$}{} theory}
\label{D*sec}
We now consider the \textbf{D}$^*$ theory, that is a mass deformation of the \textbf{D} theory in which the fundamental hypermultiplets acquires a mass. As mentioned in the Introduction, the action of the \textbf{D}$^*$  theory on a 4-sphere contains the term (\ref{massivaaction}). Without loss of generality, we can restrict the masses to be along the four Cartan directions of U(4) labeled by $i=1,\ldots,4$, {\it{i.e.}} we replace $m_A\to m_i$. There are several combinations of these masses that can occur. To find them, let us recall that the group U(4) arises from a $\mathbb{Z}_2$-orbifold projection of SO(8) (see (\ref{MZ2}) and (\ref{U4})), so that the four Cartan generators $\boldsymbol{\lambda}^i$ in the defining representation of U(4) must be embedded into $8\times 8$ matrices as
\begin{align}
    \begin{pmatrix}
        \mathbf{0}& \ii\,\boldsymbol{\lambda}^i\\
        -\ii\,\boldsymbol{\lambda}^i&\mathbf{0}
    \end{pmatrix}~.
\end{align}
We can then consider the following combination of these embedded Cartan generators
\begin{align}
    M=\begin{pmatrix}
        \mathbf{0}&\begin{matrix}
    \ii\,m_1&0&0&0\\
    0&\ii\,m_2&0&0\\
    0&0&\ii\,m_3&0\\
    0&0&0&\ii\,m_4
    \end{matrix}\\
    \begin{matrix}
    -\ii\,m_1&0&0&0\\
    0&-\ii\,m_2&0&0\\
    0&0&-\ii\,m_3&0\\
    0&0&0&-\ii\,m_4
    \end{matrix}&\mathbf{0}
    \end{pmatrix}~,
    \label{Mmatrix}
\end{align}
which satisfies
\begin{align}
    \tr M^{2k+1}=0~,~~ \tr M^{2k}=2\sum_{i=1}^{4}m_i^{2k}~,~~
    \mathrm{Pfaff}(M)= m_1\,m_2\,m_3\,m_4~.
    \label{invariants}
\end{align}
From this we see that at order 4 in the masses, there are three independent U(4)-invariant structures, which we can take to be
\begin{subequations}
    \begin{align}
    \sum_{i=1}^4 m_i^4&=\frac{1}{2}\,\tr M^4~,\label{m4}\\
    \sum_{i<j=1}^4 \!m_i^2\,m_{j}^2&=-\frac{1}{4}\,\tr M^4+\frac{1}{8}\,\big(\tr M^2\big)^2~,\label{m22}\\[1mm]
   m_1\,m_2\,m_3\,m_4&=\mathrm{Pfaff}(M)~.\label{m1111}
\end{align}
\label{massinvariants}%
\end{subequations}
Let us now turn to the matrix model of the \textbf{D}$^*$ theory. To obtain it we use the prescription of \cite{Pestun:2007rz} and make the following replacement in (\ref{Z1loop})
\begin{align}
   \prod_{i=1}^{N_F}\prod_{u=1}^N H(\ii a_u)~\to~
    \prod_{i=1}^{N_F}\prod_{u=1}^N H^{\frac{1}{2}}(\ii a_u+\ii m_i)
    \,H^{\frac{1}{2}}(\ii a_u-\ii m_i)~.
    \label{Hmass}
\end{align}
In the small-mass limit this amounts to consider a matrix model in which
\begin{align}
    \Big|Z_{\mathrm{1-loop}}\Big|^2~\to~ \Big|Z_{\mathrm{1-loop}}\Big|^2\,\,\,\rme^{\displaystyle{-\frac{1}{2}\sum_{i=1}^4m_i^2\,\sum_{u=1}^N}\partial^2\log H(\ii a_u)-\frac{1}{24}\sum_{i=1}^4m_i^4\,\sum_{u=1}^N \partial^4\log H(\ii a_u)+O(m^6)}~.
\end{align}
Notice that since only $\sum_i m_i^2$ and $\sum_i m_i^4$ appear in this expression, the resulting partition function will contain, at quartic order, only the two invariant combinations (\ref{m4}) and (\ref{m22})\,%
\footnote{This is no longer true when non-perturbative instanton contributions are taken into account.}.

Performing the rescaling (\ref{rescaling}) and then expanding for small $\lambda$, we can write the partition function of the \textbf{D}$^*$ matrix model in the following form
\begin{align}
    \cZ_{\D^*}=\int\!da\,\,\rme^{-\tr a^2 }\,\rme^{-S_\D-\sum_{i}m_i^2\,S_2-\sum_i m_i^4\,S_4+O(m^6)}
    \label{ZDstar}
\end{align}
where
\begin{align}
    S_2&=\sum_{k=1}^\infty(-1)^k\,(2k+1)\,\zeta_{2k+1}\,\Big(\frac{\lambda}{8\pi^2N}\Big)^{k}\,\tr a^{2k}~,\label{S2}\\
    S_4&=\frac{1}{6}\,\sum_{k=1}^{\infty}(-1)^k \, k\,(4k^2-1)\,\zeta_{2k+1}\Big(\frac{\lambda}{8\pi^2N}\Big)^{k-1}\,\tr a^{2k-2}~.\label{S4}
\end{align}
From this we deduce that the free energy of the \textbf{D}$^*$ theory is
\begin{align}
    \cF_{\D^*}\!
    =-\log  \cZ_{\D^*}\!
    = \cF_\D+\!\sum_{i=1}^4m_i^2\,\big\langle S_2\big\rangle_\D\!+\!\sum_{i=1}^4m_i^4\,\big\langle S_4\big\rangle_\D\!-\frac{1}{2}\Big(\sum_{i=1}^4m_i^2\Big)^2\Big(
    \big\langle S_2^{~2}\big\rangle_\D-
    \big\langle S_2\big\rangle_\D^2\Big)+O(m^6)
\end{align}
and that
\begin{subequations}
    \begin{align}
    -\partial_{m_i}^4\cF_{\D^*}\Big|_{m=0}&=-24 \,\big\langle S_4\big\rangle_\D+12\,
    \big\langle S_2^{~2}\big\rangle_\D-12\,
    \big\langle S_2\big\rangle_\D^2&& (i=1,\ldots,4)~,\label{d4f}\\[1mm]
   -\partial_{m_i}^2\partial_{m_j}^2\cF_{\D^*}\Big|_{m=0}&= 4\,
    \big\langle S_2^{~2}\big\rangle_\D-4\,
    \big\langle S_2\big\rangle_\D^2&&(i\neq j =1,\ldots,4)~.\label{d22f}
\end{align}
\label{derivatives}%
\end{subequations}
As already observed, the fourth-order mass derivative corresponding to the third U(4) invariant (\ref{m1111}) which does not appear in the matrix model, vanishes:
\begin{align}
-\partial_{m_1}\partial_{m_2}\partial_{m_3}\partial_{m_4}\cF_{\D^*}\Big|_{m=0}&=0~.
\label{d4vanish}
\end{align}
To compute the right-hand sides of (\ref{derivatives}), we rewrite $S_4$ and $S_2$ in terms of the $\cP$ operators introduced in the previous section.
Using (\ref{changebasis}) we have
\begin{align}
S_4=S_4^{(1)}+S_4^{(0)}\qquad\mbox{and}\qquad S_2=S_2^{(1)}+S_2^{(0)}
\end{align}
where the superscripts $^{(1)}$ and $^{(0)}$ denote, respectively, the part which is linear in the $\cP$ operators
and the part arising from the vacuum expectation values $\langle \tr a^{2k}\rangle_0$ which is independent of them. Explicitly we have
\begin{align}
S_4^{(1)}&=\frac{1}{12}\sum_{k=2}^\infty\sum_{\ell=0}^{k-2}(-1)^k\,(2k+1)!\,\zeta_{2k+1}\,\frac{\sqrt{2k-2-2\ell}}{\ell!\,(2k-2-\ell)!}\,
    \Big(\frac{\lambda}{16\pi^2}\Big)^{k-1}\,\cP_{2k-2-2\ell}~,\label{S41}\\
S_4^{(0)}&=\frac{1}{6}\sum_{k=1}^\infty(-1)^k\,k\,(4k^2-1)\,\zeta_{2k+1}\,\Big(\frac{\lambda}{8\pi^2N}\Big)^{k-1}\,\big\langle \tr a^{2k-2}\big\rangle_0~,\label{S40}\\
S_2^{(1)}&=\sum_{k=1}^\infty\sum_{\ell=0}^{k-1}(-1)^k\,(2k+1)!\,\zeta_{2k+1}\,\frac{\sqrt{2k-2\ell}}{\ell!\,(2k-\ell)!}\,
    \Big(\frac{\lambda}{16\pi^2}\Big)^k\,\cP_{2k-2\ell}~.\label{S21}
\end{align}
We have not written the $\cP$-independent term $S_2^{(0)}$ since in (\ref{d4f}) and (\ref{d22f}) only the connected combination $\langle S_2^{~2}\rangle_\D-\langle S_2 \rangle_\D^2$ appears and in this combination $S_2^{(0)}$ cancels out.

We notice that $S_4^{(1)}$ and $S_2^{(1)}$ have very similar expansions, which actually can be resummed in terms of Bessel functions. Indeed, using the identity
\begin{align}
(2n+1)!\,\zeta_{2n+1}= \int_0^{\infty} \frac{dt}{t}\,\frac{\rme^t}{(\rme^t-1)^2}\, t^{2n+2} ~,
\label{Zeta}
\end{align}
and defining
\begin{align}
    \mathsf{Z}_n^{(p)}=\int_0^{\infty} \frac{dt}{t}\,\frac{\rme^t\, t^{\,p}}{(\rme^t-1)^2}\,J_{n}\Big(\frac{\sqrt{\lambda}\,t}{2\pi}\Big)
    \label{Zkp}
\end{align}
for $n\geq 1$ and $p>1$,
one can check that (\ref{S41}) and (\ref{S21}) become
\begin{align}
S_4^{(1)}&=-\frac{1}{12}\,\sum_{k=1}^\infty (-1)^k
\,\sqrt{2k}\,\,\mathsf{Z}_{2k}^{(4)}\,\cP_{2k}~,\label{S4Z}\\
S_2^{(1)}&=\sum_{k=1}^\infty (-1)^k\,\sqrt{2k}\,\,\mathsf{Z}_{2k}^{(2)}\,\cP_{2k}~.
\label{S2Z}
\end{align}
Notice that, even if started from weak-coupling expansions, the final formulas (\ref{S4Z}) and (\ref{S2Z}) contain the full dependence on the 't Hooft coupling and thus are valid for any value of $\lambda$.

Also for $S_4^{(0)}$ one can find an exact expression order by order in the large-$N$ expansion. Using\,%
\footnote{This formula as well as its higher non-planar corrections can be easily derived using the fusion/fission identities of the SU($N$) traces \cite{Billo:2017glv}.}
\begin{align}
    \big\langle \tr a^{2n} \big\rangle_0 =\frac{N^{n+1}}{2^n}\,\frac{(2n)!}{n!\,(n+1)!}-\frac{N^{n-1}}{2^{n+1}}\,\frac{(2n)!}{n!\,(n-1)!}\,\Big(1-\frac{n-1}{6}\Big)+O\Big(N^{n-3}\Big)~,
\end{align}
one has
\begin{align}
    S_4^{(0)}&=\frac{N}{12}\,\sum_{k=1}^\infty (-1)^k\,(2k+1)!\,\zeta_{2k+1}\,\frac{1}{k!\,(k-1)!}\,\Big(\frac{\lambda}{16\pi^2}\Big)^{k-1}\label{S40a}\\
    &\quad-\frac{1}{144\,N}\,\sum_{k=1}^\infty (-1)^k\,(2k+1)!\,\zeta_{2k+1}\,\frac{8-k}{(k-1)!\,(k-2)!}\,\Big(\frac{\lambda}{16\pi^2}\Big)^{k-1}+O\Big(\frac{1}{N^3}\Big)~.\notag
\end{align}
Once again, after using the identity (\ref{Zeta}) one can resum the above expansions in terms of Bessel functions and show that
\begin{align}
    S_4^{(0)}=-\frac{N}{12}\,\frac{4\pi}{\sqrt{\lambda}}\,\mathsf{Z}_1^{(3)}-\frac{1}{24\,N}\,\bigg[\frac{\sqrt{\lambda}}{4\pi}\,\mathsf{Z}_1^{(5)}+\frac{1}{6}\Big(
    \,\frac{\sqrt{\lambda}}{4\pi}\Big)^2\,\mathsf{Z}_2^{(6)}\bigg]+O\Big(\frac{1}{N^3}\Big)~.
    \label{S40fin}
\end{align}

With these ingredients we can write the quantities in the right-hand side of (\ref{d4f}) and (\ref{d22f}) in closed form. Explicitly, from (\ref{S4Z}) and (\ref{S40fin}) we have
\begin{align}
 \big\langle S_4\big\rangle_\D&= -\frac{N}{12}\,\frac{4\pi}{\sqrt{\lambda}}\,\mathsf{Z}_1^{(3)}\!-\!
 \frac{1}{12}\sum_{k=1}^\infty (-1)^k\sqrt{2k}\,\,\mathsf{Z}_{2k}^{(4)}\,
 \big\langle \cP_{2k} \big\rangle_\D\!
 -\frac{1}{24\,N}\bigg[\frac{\sqrt{\lambda}}{4\pi}\,\mathsf{Z}_1^{(5)}\!+\frac{1}{6} \Big(
    \,\frac{\sqrt{\lambda}}{4\pi}\Big)^2\,\mathsf{Z}_2^{(6)}\bigg]\!+O\Big(\frac{1}{N^3}\Big)\notag\\[1mm]
    &=-\frac{N}{12}\,\frac{4\pi}{\sqrt{\lambda}}\,\mathsf{Z}_1^{(3)}-\frac{1}{12}\,\sum_{k=1}^\infty (-1)^k\,\sqrt{2k}\,\,\mathsf{Z}_{2k}^{(4)}\,\mathsf{Y}_{2k}\notag\\[1mm]
    &\qquad
    -\frac{1}{24\,N}\bigg[\frac{\sqrt{\lambda}}{4\pi}\,\mathsf{Z}_1^{(5)}+\frac{1}{6}\Big(
    \,\frac{\sqrt{\lambda}}{4\pi}\Big)^2\,\mathsf{Z}_2^{(6)}
    - \frac{\sqrt{\lambda}}{4\pi}\,\mathsf{Z}_1^{(5)}\Big({\mathsf{Y}}^{\,2}-2\lambda\,\partial_\lambda\cF_\E\Big)\bigg]+O\Big(\frac{1}{N^2}\Big)
    \label{S4fin}
\end{align}
where in the first step we have inserted the 1-point correlator (\ref{P2nD2}) and in the last term we have exploited the identity
\begin{align}
    \sum_{k=1}^\infty (-1)^k\,(2k)\,\mathsf{Z}_{2k}^{(p)}=-\frac{\sqrt{\lambda}}{4\pi}\,\mathsf{Z}_1^{(p+1)}
\end{align}
for $p=4$, which follows from the sum rule (\ref{sumBessel}). In a similar way using (\ref{S2Z}) and the connected 2-point correlator (\ref{P2nP2mc}), we find
\begin{align}
     \big\langle S_2^{\,2}\big\rangle_\D-\big\langle S_2\big\rangle_\D^2&=\sum_{k,\ell=1}^\infty
     (-1)^{k+\ell}\,\sqrt{2k}\,\sqrt{2\ell}\,\,\mathsf{Z}_{2k}^{(2)}\,\,\mathsf{Z}_{2\ell}^{(2)}\,\,\big\langle \cP_{2k}\,\cP_{2\ell}\big\rangle_\D^c\notag\\[1mm]
     &=\sum_{k=1}^\infty 2k\,\big(\mathsf{Z}_{2k}^{(2)}\big)^2+
     \frac{1}{N}\Big(\frac{\sqrt{\lambda}}{4\pi}\,\,\mathsf{Z}_1^{(3)}\Big)^2\,{\mathsf{Y}}+O\Big(\frac{1}{N^2}\Big)~.
\label{S2fin}
\end{align}
Both (\ref{S4fin}) and (\ref{S2fin}) are examples of large-$N$ expansions in which, order by order in $1/N$, the dependence on $\lambda$ is exact and given through integrals of Bessel functions. It is worth underlining that the procedure we have described can be used also to derive further sub-leading contributions. Even if the calculations soon become quite challenging, in principle there are no obstacles to extend our results to higher orders in $1/N$ since everything is purely algebraic.

\section{Strong coupling limit}
\label{secn:strong}
The results of the previous section allow us to study the mass-derivatives (\ref{derivatives}) as functions of the 't Hooft coupling. When $\lambda$ is small, we can Taylor-expand the Bessel functions inside the various integrals and obtain the following weak-coupling expansions
\begin{subequations}
    \begin{align}
    -\partial_{m_i}^4\cF_{\D^*}\Big|_{m=0}~ &\underset{\lambda \rightarrow 0}{\sim}~N\big(12\,\zeta_3-60\,\zeta_5\,\hat\lambda+210\,\zeta_7\,\hat\lambda^2-630\,\zeta_9\,\hat\lambda^3+\cdots\big)\notag\\
    &\quad\quad+
    \big(54\,\zeta_3^2\,\hat\lambda^2+360\,\zeta_3\,\zeta_5\,\hat\lambda^3-4725\,\zeta_3\,\zeta_7\,\hat\lambda^4-3825\,\zeta_5^2\,\hat\lambda^4+\cdots\big)\notag\\[1mm]
    &\quad\quad+\frac{1}{N} \big(60\,\zeta_5\,\hat\lambda-525\,\zeta_7\,\hat\lambda^2+2520\,\zeta_9\,\hat\lambda^3+\cdots\big)+O\Big(\frac{1}{N^2}\Big)~,\label{d4Fweak}\\[2mm]
   -\partial_{m_i}^2\partial_{m_j}^2\cF_{\D^*}\Big|_{m=0}~ &\underset{\lambda \rightarrow 0}{\sim}~\big(18\,\zeta_3^2\,\hat\lambda^2-120\,\zeta_3\,\zeta_5\,\hat\lambda^3+\cdots\big)\notag\\
    &\quad\quad+\frac{1}{N} \big(\!-648\,\zeta_3^3\,\hat\lambda^4+11880\,\zeta_3^2\,\zeta_5\,\hat\lambda^5+\cdots\big)+O\Big(\frac{1}{N^2}\Big)\label{d22Fweak}
\end{align}
\label{derivativesweak}%
\end{subequations}
where we have defined $\hat\lambda=\lambda/(8\pi^2)$ for ease of notation.
The perturbative series in round brackets can be computed to very high order with a limited computational effort and can be shown to have a finite radius of convergence ($|\lambda|\leq \pi^2$). By using standard resummation methods one can extend them beyond this limit and explore the strong-coupling regime. Actually, since we have derived explicit expressions in terms of integrals of Bessel functions that are valid for any $\lambda$, we do not have to follow this path and we can simply exploit the asymptotic behavior of the Bessel functions combined with the Mellin-Barnes method to obtain the asymptotic expansions for large values of $\lambda$. This procedure is described in detail in Appendix~\ref{app:strong} and here we just report the results that are needed to evaluate the right-hand sides of (\ref{d4f}) and (\ref{d22f}). More specifically, from (\ref{strongZkp}) and (\ref{Zstrong}) we read that
\begin{align}
   \mathsf{Z}_1^{(3)} \underset{\lambda \rightarrow \infty}{\sim} ~
   \frac{2\pi}{\sqrt{\lambda}}+O\big(\rme^{-\lambda}\big)~,\quad
   \mathsf{Z}_1^{(5)} \underset{\lambda \rightarrow \infty}{\sim} ~
   0+O\big(\rme^{-\lambda}\big)~,\quad
   \mathsf{Z}_2^{(6)} \underset{\lambda \rightarrow \infty}{\sim} ~
  0+O\big(\rme^{-\lambda}\big)~,
  \label{Zasym}
\end{align}
from (\ref{Ystrongapp}) that
\begin{align}
   \mathsf{Y} \underset{\lambda \rightarrow \infty}{\sim} -\frac{\log2}{2\pi^2}\,\lambda+\frac{1}{4}+O\big(\rme^{-\lambda}\big)~,
   \label{Yasym}
\end{align}
from (\ref{ZZstrongapp}) that
\begin{align}
    \sum_{k=1}^\infty 2k\,\big(\mathsf{Z}_{2k}^{(2)}\big)^2~ \underset{\lambda \rightarrow \infty}{\sim} ~
    \frac{1}{4}\log\lambda + \frac{1}{2}\,\gamma - \frac{1}{2}\log(4\pi)-\frac{1}{2}\,\zeta_3+\frac{11}{12}+O\big(\rme^{-\lambda}\big)~,
    \label{ZZasym}
\end{align}
and finally from (\ref{ZYstrongapp}) that
\begin{align}
    \sum_{k=1}^\infty (-1)^k\,\sqrt{2k}\,\,\mathsf{Z}_{2k}^{(4)}\,\mathsf{Y}_{2k}~ \underset{\lambda \rightarrow \infty}{\sim} ~
    \frac{3}{2}\,\zeta_3+O\big(\rme^{-\lambda}\big)~.
    \label{ZYasym}
\end{align}
Notice that in all cases, apart from the non-perturbative exponentially suppressed terms, the asymptotic series contain only a finite number of terms and in some cases, like for example in $\mathsf{Z}_1^{(5)}$, none. This is a known phenomenon which occurs when some parameters take special values, typically integers, at which the usual asymptotic expansion terminates or even disappears. In these cases the tail of non-perturbative exponentially small corrections is obtained using the so-called Cheshire cat resurgence (see for instance \cite{Dorigoni:2017smz} and references therein).

The fact that the $\mathsf{Z}_1^{(5)}$ asymptotically vanishes causes a big simplification since the only term at order $1/N$ in (\ref{S4fin}) that depends on the free energy of the \textbf{E} theory which has a strong-coupling expansion with infinite terms \cite{Beccaria:2021vuc,Beccaria:2022ypy}, drops out being multiplied by $\mathsf{Z}_1^{(5)}$. Because of this fact, at each order in the $1/N$ expansion, only a finite number of $\lambda$-dependent terms appear at strong coupling. Indeed,
inserting in (\ref{d4f}) and (\ref{d22f}) the results (\ref{Zasym})-(\ref{ZYasym}), after simple algebra we find
\begin{subequations}
\begin{align}
-\partial_{m_i}^4\cF_{\D^*}\Big|_{m=0} ~  & \underset{\lambda \rightarrow \infty}{\sim} ~ \frac{16\pi^2}{\lambda}N +3\log\lambda +6\gamma - 6\log(4\pi) -3\,\zeta_3 +11
\notag \\[1mm] & 
\qquad\qquad+ \frac{3}{4 N}\Big(1-\frac{2\log 2 }{\pi^2}\,\lambda \Big)+
O\Big(\frac{1}{N^2}\Big)~, \label{d4Fstrong}\\[2mm]
-\partial_{m_i}^2\partial_{m_j}^2\cF_{\D^*}\Big|_{m=0}~& \underset{\lambda \rightarrow \infty}{\sim} ~ \log\lambda +2\gamma -2\log (4\pi)-2\,\zeta_3+\frac{11}{3} \nonumber\\
&\qquad\qquad +\frac{1}{4N}\Big(1-\frac{2\log 2}{\pi^2}\,\lambda\Big) +O\Big(\frac{1}{N^2}\Big) ~.
\label{d22Fstrong}
\end{align}
\end{subequations}%
The $\log 2$ terms in the round brackets of the above formulas can be removed by introducing a shifted 't Hooft coupling defined according to
\begin{align}
    \frac{1}{\lambda^\prime}=\frac{1}{\lambda}+\frac{\log 2}{2\pi^2N}~.
    \label{lambdaprime}
\end{align}
In terms of $\lambda^\prime$ we have
\begin{subequations}
\begin{align}
-\partial_{m_i}^4\cF_{\D^*}\Big|_{m=0} ~  & \underset{\lambda^\prime \rightarrow \infty}{\sim} ~ \frac{16\pi^2}{\lambda^\prime}N +3\log\lambda^\prime+3f(N) -8\log 2 +3\,\zeta_3~, \label{d4Fstrong1}\\[2mm]
-\partial_{m_i}^2\partial_{m_j}^2\cF_{\D^*}\Big|_{m=0}~& \underset{\lambda^\prime \rightarrow \infty}{\sim} ~ \log\lambda^\prime +f(N)
\label{d22Fstrong1}
\end{align}
\label{d4d22strong}
\end{subequations}%
where
\begin{align}
    f(N)=2\gamma -2\log (4\pi)-2\,\zeta_3+\frac{11}{3}+\frac{1}{4N}+O\Big(\frac{1}{N^2}\Big)~.
\end{align}
The strong-coupling formulas (\ref{d4Fstrong1}) and (\ref{d22Fstrong1}) are one of our main results. They are very similar (but not identical) to the findings of \cite{Behan:2023fqq}
where the $\cN=2$ Sp($N$) SYM theory with one anti-symmetric and four fundamental hypermultiplets has been analyzed. In this theory the flavor group of the fundamental matter is SO(8), instead of U(4), but the quartic invariants $\sum_i m_i^4$, $\sum_{i<j} m_i^2m_j^2$ and $(m_1\,m_2\,m_3\,m_4)$ are the same, and thus the corresponding quartic derivatives of the partition function can be compared in the two theories. Confronting Eq.\,(3.13) of \cite{Behan:2023fqq} with our equations (\ref{d4d22strong}) and (\ref{d4vanish}), we obtain
\begin{subequations}
\begin{align}
-\partial_{m_i}^4\cF_{\D^*}\Big|_{m=0} &=\frac{1}{2}\bigg[-\partial_{m_i}^4\cF_{\mathrm{Sp}^*}\Big|_{m=0}\bigg]+3\,\zeta_3-\frac{3}{4N}+O\Big(\frac{1}{N^2}\Big)~, \label{d4FSuSp}\\[2mm]
-\partial_{m_i}^2\partial_{m_j}^2\cF_{\D^*}\Big|_{m=0}&=\frac{1}{2}\bigg[-\partial_{m_i}^2\partial_{m_j}^2\cF_{\mathrm{Sp}^*}\Big|_{m=0}\bigg]-\frac{1}{4N}+O\Big(\frac{1}{N^2}\Big)
\label{d22FSuSp}\\[2mm]
-\partial_{m_1}\partial_{m_2}\partial_{m_3}\partial_{m_4}
\cF_{\D^*}\Big|_{m=0}&=\frac{1}{2}\bigg[-\partial_{m_1}\partial_{m_2}\partial_{m_3}\partial_{m_4}\cF_{\mathrm{Sp}^*}\Big|_{m=0}\bigg]+O\Big(\frac{1}{N^2}\Big)
\label{d1111FSuSp}
\end{align}
\label{dSpvsU}%
\end{subequations}
where we have denoted by $\cF_{\mathrm{Sp}^*}$ the free energy of the Sp($N$) theory in which the fundamental hypermultiplets are massive. These relations show that the \textbf{D} and the Sp($N$) theories are planar equivalent at strong coupling (up to a trivial overall factor of 1/2) but differ at the non-planar level by a function which, up to the order $1/N^2$, depends only on $N$. This fact provides strong evidence that the same thing happens also to higher orders in the $1/N$-expansion, suggesting that in the strong-coupling regime the dependence on the 't Hooft coupling is the same in the two theories, apart from an overall factor. In other words, this argument supports that (\ref{d4d22strong}) capture the full dependence on $\lambda^\prime$ at strong coupling and large $N$, with $f(N)$ remaining a function only of $N$ also at higher orders, in complete analogy with the Sp($N$) theory \cite{Behan:2023fqq}.

We finally observe that more than a redefinition of the 't Hooft coupling, the shift (\ref{lambdaprime}) is actually a redefinition of the Yang-Mills coupling which can be written as
\begin{align}
    \frac{8\pi^2}{g^{\prime\,2}}=\frac{8\pi^2}{g^2}+4\log 2~.
    \label{gprime}
\end{align}
or, introducing the complex combination
\begin{align}
   \tau=\frac{\theta}{2\pi}+\ii\,\frac{4\pi}{g^2}\,\equiv\,\tau_1+\ii\,\tau_2 
\end{align}
where $\theta$ is the vacuum $\theta$-angle, as
\begin{align}
    q=16\,q^\prime
    \label{qprime}
\end{align}
with $q=\rme^{2\pi\ii\tau}$ and\,\footnote{We follow the standard normalizations and conventions of the literature for the IR complex coupling $\tau^\prime$; see for example \cite{Seiberg:1994aj}.}
\begin{align}
    q^\prime=\rme^{\ii\pi\tau^\prime}~,~~~ \tau^\prime=\frac{\theta^\prime}{\pi}+\ii\,\frac{8\pi}{g^{\prime\,2}}~.
    \label{qprime0}
\end{align}
It is interesting to observe that this is the same as the UV/IR relation of SQCD with gauge group SU(2)\,%
\footnote{Notice that for SU(2) the anti-symmetric hypermultiplet is a singlet, which decouples, and only the four fundamental hypermultiplets remain. Thus, for SU(2) the \textbf{D} theory effectively coincides with SQCD.}. More precisely, in this case one has \cite{Alday:2009aq} (see also \cite{Billo:2010mg} and references therein)
\begin{align}
q=\frac{\theta_2(q^\prime)^4}{\theta_3(q^\prime)^4}
\label{qqprimetheta}
\end{align}
where $\theta_2$ and $\theta_3$ are Jacobi $\theta$-functions. Expanding the right-hand side in powers of $q^\prime$, at the leading order one recovers precisely the relation (\ref{qprime}), with the sub-leading terms representing non-perturbative instanton corrections. This fact suggests that the \textbf{D} theory at large $N$
might share the same non-perturbative and modular structure as the SU(2) theory. 
Note that this is strictly analogous to the situation that occurs in the $\cN=2$ superconformal Sp(1) theory with four fundamental hypermultiplets for which it has been established \cite{Hollands:2010xa} that
\begin{align}
q^2=16\,\frac{\theta_2(q^{\prime\,2})^4}{\theta_3(q^{\prime\,2})^4}~.
\label{qqprimethetaSp}
\end{align}
Again neglecting instantons, this leads to the same relation (\ref{qprime}) and hence to (\ref{lambdaprime}) and (\ref{gprime}). In \cite{Beccaria:2022kxy} this relation has been observed also in the superconformal Sp($N$) theory in the large-$N$ limit at strong coupling, and in \cite{Behan:2023fqq} it has been promoted at the full non-perturbative level, thus extending the Sp(1) formula (\ref{qqprimethetaSp}) to the Sp($N$) theory at large $N$. It would be very interesting to check whether or not this non-perturbative completion occurs also in our \textbf{D} theory.

Finally, we observe that the simple relationship between the strong-coupling behavior of the mass-derivatives in the Sp($N$) and \textbf{D} theories shown in (\ref{dSpvsU}) suggests that the same calculations performed in \cite{Alday:2024yax,Behan:2024vwg,Alday:2024ksp} to fix the coefficients of the AdS scattering amplitude of four SO(8) gluons from localization can also be repeated for the scattering of U(4) gluons, leading to the same results. Our matrix-model findings in the \textbf{D} theory are therefore useful for constraining the dual gluon amplitudes in AdS\,%
\footnote{We thank Tobias Hansen for correspondence on this point.}.

\section{Conclusions and outlook}
\label{sec:concl}
Using the full Lie algebra approach to matrix models we have studied in detail the derivatives of the free energy of the \textbf{D}$^*$ theory order by order in the large-$N$ expansion, obtaining exact expressions that are valid for all values of the 't Hooft coupling. In our analysis we have worked up to the first three non-trivial orders in the $1/N$ expansion ({\it{i.e.}} up to order $1/N$) but in principle there are no obstacles to going further. It would be interesting to explore the possibility of finding a systematic way to compute and organize these higher order corrections, similarly to what happens for the matrix model of the Sp($N$). Here one can exploit a Toda lattice equation \cite{Beccaria:2021ism,Beccaria:2022kxy} and establish recursive relations among theories with $N$ and $N\pm1$, which then allow obtaining formulas that are exact in $N$. To the best of our knowledge, however, this Toda lattice equation cannot be applied to matrix models with double-trace interactions like that of the \textbf{D} theory; so this remains an open problem.  

In this paper we have considered a deformation of the \textbf{D} theory in which the four fundamental hypermultiplets acquire a mass. One could have given a mass also to the two anti-symmetric hypermultiplets, and then consider mixed quartic derivatives with respect to the two types of masses. Such derivatives would correspond to integrated correlators of four moment map operators, two of which are dual to gluons and two to closed string excitations of the bulk sectors. Thus, at strong coupling in the large-$N$ limit these mixed quartic derivatives could provide information about mixed open/closed string amplitudes in AdS. As far as we know, this line of investigation has not yet been pursued so far and it would be interesting to begin exploring it.

As we have emphasized several times, the matrix model approach we have described yields, order by order in the $1/N$ expansion, explicit formulas that are written in terms of integrals of Bessel functions and are valid for all values of the coupling constant. Exploiting the asymptotic properties of the Bessel functions, we have studied the behavior at large values of the 't Hooft coupling finding in all cases examined that the asymptotic expansion actually ends after a finite number of terms or even does not exist. It would therefore be interesting to apply the Cheshire cat resurgence methods \cite{Dorigoni:2017smz} to determine the non-perturbative exponentially suppressed corrections $O(\rme^{-\sqrt{\lambda}})$ and investigate their interpretation in the holographic dual theory.

Finally, it would be important to study the large-$N$ limit at fixed Yang-Mills coupling. In this regime the instanton corrections cannot be neglected and produce non-perturbative terms proportional to powers of $q^\prime$ and $\overbar{q}^\prime$ which must be added to the perturbative results we have derived in this paper. It would be very interesting to compute these terms and check whether, as expected, they provide the completion of the present results into modular functions. This analysis would therefore shed light on the modular properties of the \textbf{D} theory in the large $N$ regime. Work along these lines is in progress.

\vskip 1cm
\noindent {\large {\bf Acknowledgments}}
\vskip 0.2cm
We would like to thank Daniele Dorigoni, Pietro Ferrero, Francesco Galvagno, Tobias Hansen, Gregory Korchemsky, Oliver Schlotterer, Yifan Wang and Congkao Wen for many useful discussions. We all thank the Galileo Galilei Institute (GGI) in Florence for hospitality and support during the workshop ``Resurgence and Modularity in QFT and String Theory''. MB thanks the RIKEN Interdisciplinary Theoretical and Mathematical Sciences Program and the Yukawa Institute for Theoretical Physics (YITP) at Kyoto University. Discussions during the ``iTHEMS-YITP workshop: Bootstrap, Localization and Holography'' were useful to complete this work.

This research is partially supported by the MUR PRIN contract 2020KR4KN2 ``String Theory as a bridge between Gauge Theories and Quantum Gravity'' and by the INFN project ST\&FI ``String Theory \& Fundamental Interactions''. The work of AP is supported  by the Deutsche Forschungsgemeinschaft (DFG, German Research Foundation) via the Emmy Noether program ``Exploring the landscape of string theory flux vacua using exceptional field theory” (project number 426510644). 

\vskip 1cm

\appendix

\section{Details on the strong coupling expansion}
\label{app:strong}

In this appendix we provide some technical details on the derivation of the strong-coupling expansion of the quantities considered in the main text. We analyse them one by one. 

\subsubsection*{Strong coupling behavior of \texorpdfstring{$\mathsf{Z}_n^{(p)}$}{}}
$\mathsf{Z}_n^{(p)}$ is defined in \eqref{Zkp} which we rewrite here for convenience:
\begin{align}
    \mathsf{Z}_n^{(p)}=\int_0^{\infty} \frac{dt}{t}\,\frac{\rme^t\, t^{\,p}}{(\rme^t-1)^2}\,J_{n}\Big(\frac{\sqrt{\lambda}\,t}{2\pi}\Big)
\end{align}
for $n\geq 1$ and $p>1$.
In order to study its strong-coupling expansion, we use the Mellin-Barnes integral representation of the Bessel function
\begin{align}
\label{Jmellin}
J_{n}(x) =  \int_{-\ii \infty}^{+\ii \infty} \frac{ds}{2 \pi \ii}\,\frac{\Gamma(-s)}{\Gamma(s+n+1)}\,\Big(\frac{x}{2}\Big)^{2s+n}~,
\end{align} 
and obtain
\begin{align}
    \mathsf{Z}_n^{(p)}=\int_0^{\infty} \frac{dt}{t}\,\frac{\rme^t\, t^{\,p}}{(\rme^t-1)^2}\, \int_{-\ii \infty}^{+\ii \infty} \frac{ds}{2 \pi \ii}\,\frac{\Gamma(-s)}{\Gamma(s+n+1)}\,\Big(\frac{\sqrt{\lambda}\,t}{4\pi}\Big)^{2s+n}~.
\end{align}
Evaluating the $t$-integral using \eqref{Zeta}, we get
\begin{align}
    \mathsf{Z}_n^{(p)}=\int_{-\ii \infty}^{+\ii \infty} \frac{ds}{2 \pi \ii}\,\frac{\Gamma(-s)\,\Gamma(2s+n+p)\,\zeta_{2s+n+p-1}}{\Gamma(s+n+1)}\,\Big(\frac{\sqrt{\lambda}}{4\pi}\Big)^{2s+n}~.
\end{align}
When $\lambda\to\infty$ this integral receives contributions from poles on the negative real axis of $s$. Summing the residues over such poles, one finds
\begin{align}
    \mathsf{Z}_n^{(p)} \underset{\lambda \rightarrow \infty}{\sim}~ -\frac{1}{2}
    \sum_{k=0}^\infty \frac{(2k-1)\,B_{2k}}{(2k)!}\,\frac{\Gamma\big(\frac{n+p}{2}+k-1\big)}{\Gamma\big(\frac{n-p}{2}+2-k\big)}\,
    \Big(\frac{4\pi}{\sqrt{\lambda}}\Big)^{p+2k-2}
    \label{strongZkp}
\end{align}
where $B_{2k}$ are the Bernoulli numbers.
From this, it is straightforward to deduce that
\begin{align}
 \mathsf{Z}_1^{(3)} \underset{\lambda \rightarrow \infty}{\sim}~ \frac{2\pi}{\sqrt{\lambda}}\,, \qquad \mathsf{Z}_1^{(5)} \underset{\lambda \rightarrow \infty}{\sim}~ 0\,, \qquad \mathsf{Z}_2^{(6)} \underset{\lambda \rightarrow \infty}{\sim}~ 0~. 
 \label{Zstrong}
\end{align}
We observe that when $n$ and $p$ are both even or both odd, the asymptotic expansion (\ref{strongZkp}) terminates after a finite number of terms or even disappears as for example in $\mathsf{Z}_1^{(5)}$ or $\mathsf{Z}_{2}^{(6)}$. This fact is not in contradiction with the fact that $\mathsf{Z}_n^{(p)}$ is always a non-trivial function of $\lambda$. Indeed, beside the asymptotic series (\ref{strongZkp}) one should consider also non-perturbative exponentially suppressed terms that are obtained by using the so-called Cheshire cat resurgence (see for instance \cite{Dorigoni:2017smz} and references therein).

\subsubsection*{Strong coupling behavior of \texorpdfstring{$\mathsf{Y}$}{}}
The quantity $\mathsf{Y}$ is defined \eqref{Yhat}, namely
\begin{align}
    {\mathsf{Y}} =
    \int_0^\infty \frac{dt}{t}\,\frac{\rme^t}{(\rme^t+1)^2}\,
\bigg[\frac{\sqrt{\lambda}\,t}{\pi}\,J_{1}\Big(\frac{\sqrt{\lambda}\,t}{\pi}\Big)\bigg]-\frac{\log 2}{2\pi^2}\,\lambda~.
\end{align}
After using \eqref{Jmellin} and (\ref{zetaidentity}), we get
\begin{align}
    {\mathsf{Y}} &=
    2\int_0^\infty \frac{dt}{t}\,\frac{\rme^t}{(\rme^t+1)^2}\,\int_{-\ii \infty}^{+\ii \infty} \frac{ds}{2 \pi \ii}\,\frac{\Gamma(-s)}{\Gamma(s+2)} \Big(\frac{\sqrt{\lambda}\,t}{2\pi} \Big)^{2s+2} -\frac{\log 2}{2\pi^2}\,\lambda\notag\\[1mm]
    &=
    2 \int_{-\ii \infty}^{+\ii \infty} \frac{ds}{2 \pi \ii}\,\frac{\Gamma(-s)\,\Gamma(2s+2)\,\eta_{2s+1}}{\Gamma(s+2)} \Big(\frac{\sqrt{\lambda}}{2\pi} \Big)^{2s+2} -\frac{\log 2}{2\pi^2}\,\lambda~.
\end{align}
Picking up the residue at $s=-1$, which is the only pole located on the negative real axis, we find
\begin{align}
 \mathsf{Y} \underset{\lambda \rightarrow \infty}{\sim}~ -\frac{\log 2}{2\pi^2}\,\lambda +\frac{1}{4} 
 \label{Ystrongapp}
\end{align}
up to exponentially suppressed non-perturbative terms.

\subsubsection*{Strong coupling behavior of \texorpdfstring{$\displaystyle{\sum_{k=1}^\infty 2k\,\big(\mathsf{Z}_{2k}^{(2)}\big)^2}$}{}}

Using the Mellin-Barnes representation \eqref{Jmellin} in the definition \eqref{Zkp}, we obtain
\begin{align}
\sum_{k=1}^\infty 2k\,\big(\mathsf{Z}_{2k}^{(2)}\big)^2
=&\sum_{k=1}^{\infty}2k  \int_{0}^{\infty} \frac{dt}{t}\, \frac{\rme^t\,t^2}{(\rme^t-1)^2}\int_{0}^{\infty} \frac{dt^\prime}{t^\prime} \, \frac{\rme^{t^\prime}\,t^{\prime\,2}}{(\rme^{t^\prime}-1)^2}\int_{-\ii \infty}^{+\ii \infty} \frac{ds}{2 \pi \ii}\int_{-\ii \infty}^{+\ii \infty} \frac{ds^\prime}{2 \pi \ii} \notag \\[1mm]
&\quad \times ~\frac{\Gamma(-s)\,\Gamma(-s^\prime)}{\Gamma(s+2k+1)\,\Gamma(s^\prime+2k+1)}\,
t^{2s+2k}\,t^{\prime\,2s^\prime+2k}\,
\Big(\frac{\sqrt{\lambda}}{4\pi} \Big)^{2s+2s^\prime+4k} ~.
\end{align}
Evaluating the integrals over $t$ and $t^\prime$ we find
\begin{align}
\sum_{k=1}^\infty 2k\,\big(\mathsf{Z}_{2k}^{(2)}\big)^2 = & \sum_{k=1}^{\infty}2k \int_{-\ii \infty}^{+\ii \infty} \frac{ds}{2 \pi \ii}\int_{-\ii \infty}^{+\ii \infty} \frac{ds^\prime}{2 \pi \ii}\, \Big(\frac{\sqrt{\lambda}}{4\pi} \Big)^{2s+2s^\prime+4k}\,\,\zeta_{2s+2k+1}\,\zeta_{2s'+2k+1} \notag \\[1mm]
& \quad\times~\frac{\Gamma(-s)\,\Gamma(-s^\prime)\,\Gamma(2s+2k+2)\,\Gamma(2s^\prime+2k+2)}{\Gamma(s+2k+1)\,\Gamma(s^\prime+2k+1)}\notag\\[1mm]
=& \sum_{k=1}^{\infty}2k \int_{-\ii \infty}^{+\ii \infty} \frac{ds}{2 \pi \ii}\int_{-\ii \infty}^{+\ii \infty} \frac{ds^\prime}{2 \pi \ii} \,\Big(\frac{\sqrt{\lambda}}{4\pi} \Big)^{2s+2s^\prime+4}\,\,\zeta_{2s+3}\,\zeta_{2s^\prime+3} \notag \\ 
&\quad \times~ \frac{\Gamma(k-s-1)\,\Gamma(k-s^\prime-1)\,\Gamma(2s+4)\,\Gamma(2s^\prime+4)}{\Gamma(s+k+2)\,\Gamma(s^\prime+k+2)}
\label{Z2kapp}
\end{align}
where the last step follows from shifting $s\mapsto s+1-k$ and $s^\prime \mapsto s^\prime+1-k$. We now perform the sum over $k$ by means of the identity
\begin{align}
    \sum_{k=1}^\infty k\,\,\frac{\Gamma(k-s-1)\,\Gamma(k-s^\prime-1)}{\Gamma(k+s+2)\,\Gamma(k+s^\prime+2)}\,=\,\frac{1}{2(s+s^\prime+2)}\,\frac{\Gamma(-s)\,\Gamma(-s^\prime)}{\Gamma(s+2)\,\Gamma(s^\prime+2)}~,
    \label{identity}
\end{align}
and get
\begin{align}
\sum_{k=1}^\infty 2k\,\big(\mathsf{Z}_{2k}^{(2)}\big)^2 = &
\int_{-\ii \infty}^{+\ii \infty} \frac{ds}{2 \pi \ii}\int_{-\ii \infty}^{+\ii \infty} \frac{ds^\prime}{2 \pi \ii} \,\Big(\frac{\sqrt{\lambda}}{4\pi} \Big)^{2s+2s^\prime+4}\,\,\zeta_{2s+3}\,\zeta_{2s^\prime+3}\notag \\ 
&\quad \times~ \frac{\Gamma(-s)\,\Gamma(-s^\prime)\,\Gamma(2s+4)\,\Gamma(2s^\prime+4)}{(s+s^\prime+2)\,\Gamma(s+2)\,\Gamma(s^\prime+2)}~.
\label{ZZapp}
\end{align}
When we close the integration contours in an anti-clockwise way, we
pick up two types of contributions: the one from the residues at $(s=-1\,,\,s^\prime=-1)$, and the one from the residues at $(s=-s^\prime-2\,,\,s^\prime=-n)$ for
$n=1,2,\cdots$. The first contribution is
\begin{align}
    \frac{1}{4}\log\lambda + \frac{1}{2}\gamma - \frac{1}{2}\log(4\pi)+\frac{1}{2}~,
    \label{I}
\end{align}
while the second is formally
\begin{align}
    -\frac{1}{2}\sum_{n=1}^\infty B_{2n}\,(4n^2-1)\,\zeta_{2n+1}~.
    \label{II}
\end{align}
As it stands, this sum is divergent but we can easily regularize it
by introducing the integral representation (\ref{Zeta}) of the $\zeta$-values and then using the generating function of the even Bernoulli numbers, namely
\begin{align}
   f(t)\,\equiv\, \sum_{n=1}^\infty \frac{B_{2n}}{(2n)!} \,t^{2n}=\frac{t}{\rme^t-1}-1+\frac{t}{2}~.
    \label{Bernoulligen}
\end{align}
Doing this, the sum (\ref{II}) can be rewritten as
\begin{align}
-\frac{1}{2}\sum_{n=1}^\infty B_{2n}\,(4n^2-1)\,\zeta_{2n+1}
&=
    -\frac{1}{2}\int_0^\infty
    \frac{dt}{t}\,\frac{\rme^t\,t^2}{(\rme^t-1)^2} \sum_{n=1}^\infty \frac{(2n-1)\,B_{2n}}{(2n)!}\,t^{2n}\notag\\
    &=
    -\frac{1}{2}\int_0^\infty
    \frac{dt}{t}\,\frac{\rme^t\,t^2}{(\rme^t-1)^2} \,\Big[t\,\partial_t f(t)-f(t)\Big]
    =\frac{5}{12}-\frac{1}{2}\,\zeta_3~.
    \label{IIa}
\end{align}
Adding (\ref{I}) and (\ref{IIa}), we obtain
\begin{align}
\sum_{k=1}^\infty 2k\,\big(\mathsf{Z}_{2k}^{(2)}\big)^2 \underset{\lambda \rightarrow \infty}{\sim} ~\frac{1}{4}\log\lambda + \frac{1}{2}\gamma - \frac{1}{2}\log(4\pi)-\frac{1}{2}\zeta_3+\frac{11}{12} ~.
\label{ZZstrongapp}
\end{align}
Notice that this result agrees with the one derived in Appendix A of \cite{Behan:2023fqq} with different methods. Indeed one can check that the expression in the right-hand side of (\ref{ZZapp}) equals $-\frac{F^{(2,2)}}{16}$, where $F^{(2,2)}_2$ is defined in (A.26) of \cite{Behan:2023fqq}. Using this information, we readily obtain 
(\ref{ZZstrongapp}).

\subsubsection*{Strong coupling behavior of \texorpdfstring{$\displaystyle{\sum_{k=1}^{\infty}(-1)^k\sqrt{2k}\,\mathsf{Z}_{2k}^{(4)}\,\mathsf{Y}_{2k}}$}{}}

Combining the definitions \eqref{Zkp} and \eqref{Yvector} and the Mellin-Barnes representation \eqref{Jmellin}, we can write
\begin{align}
    \sum_{k=1}^{\infty}(-1)^k\sqrt{2k}\,\mathsf{Z}_{2k}^{(4)}\,\mathsf{Y}_{2k}= \cA+\cB
    \label{ZYappAB}
\end{align}
where
\begin{align}
    \cA &=\frac{\log 2}{2\pi^2}\,\lambda \int_{0}^{\infty} \frac{dt}{t}\, \frac{\rme^{t}\,t^4}{(\rme^t-1)^2} \,\int_{-\ii \infty}^{+\ii \infty} \frac{ds}{2 \pi \ii}\,\frac{\Gamma(-s)}{\Gamma(s+3)}\, \Big(\frac{\sqrt \lambda\,t}{4 \pi}\Big)^{2s+2} \notag\\
    &=\frac{\log 2}{2\pi^2}\,\lambda \int_{-\ii \infty}^{+\ii \infty} \frac{ds}{2 \pi \ii}\,\frac{\Gamma(-s)\,\Gamma(2s+6)\,\zeta_{2s+5}}{\Gamma(s+3)}\, \Big(\frac{\sqrt \lambda}{4 \pi}\Big)^{2s+2}
    \label{Aapp}
\end{align}
and
\begin{align}
    \cB&=- \sum_{k=1}^{\infty} 4k  \int_0^\infty \frac{dt}{t}\,\frac{\rme^{t}}{(\rme^{t}+1)^2} \int_{0}^{\infty} \frac{dt^\prime}{t^\prime}\, \frac{\rme^{t^\prime}\,t^{\prime\,4}}{(\rme^{t^\prime}-1)^2} \int_{-\ii \infty}^{+\ii \infty} \frac{ds}{2 \pi \ii} \int_{-\ii \infty}^{+\ii \infty} \frac{ds^\prime}{2 \pi \ii}
    \notag \\[1mm]
&  \qquad \times ~\Big(\frac{\sqrt{\lambda} \,t }{2\pi}\Big)^{2s+2k}\,\Big(\frac{\sqrt{\lambda} \,t }{4\pi}\Big)^{2s^\prime+2k}\,\frac{\Gamma(-s)\,\Gamma(-s')}{\Gamma(s+2k+1)\,\Gamma(s'+2k+1)}\notag\\[1mm]
&=- \sum_{k=1}^\infty k \int_{-\ii \infty}^{+\ii \infty} \frac{ds}{2 \pi \ii} \int_{-\ii \infty}^{+\ii \infty} \frac{ds^\prime}{2 \pi \ii} \,\Big( \frac{\sqrt{\lambda}}{\pi} \Big)^{2s+2s^\prime+4k} \,\Big( \frac{1}{2} \Big)^{2s+4s^\prime+6k-2} \notag \\[1mm]
&\qquad \times ~ \frac{\Gamma(-s)\,\Gamma(2s+2k)}{\Gamma(s+2k+1)}\,
 \eta_{2s+2k-1} \,\,\frac{\Gamma(-s')\,\Gamma(2s'+2k+4)}{\Gamma(s'+2k+1)}\,\zeta_{2s'+2k+3}~.
 \label{sumZYint}
\end{align}
Here we have performed the integrals over $t$ and $t^\prime$ using
(\ref{zetaidentity}) and (\ref{Zeta}).

The strong-coupling behavior of $\cA$ can be easily obtained by closing the contour in an anti-clockwise manner and picking up the
residue at the only pole located at $s=-2$. This yields
\begin{align}
    \cA \underset{\lambda \rightarrow \infty}{\sim} \,4\log 2 ~.
    \label{Astrong}
\end{align}
To obtain the strong-coupling behavior of $\cB$, as in the previous case, we first perform the shifts $s\mapsto s+1-k$ and $s' \mapsto s'+1-k$, and then sum over $k$ exploiting the identity (\ref{identity}). Proceeding in this way, we get
\begin{align}
    \cB&=- \sum_{k=1}^\infty k \int_{-\ii \infty}^{+\ii \infty} \frac{ds}{2 \pi \ii} \int_{-\ii \infty}^{+\ii \infty} \frac{ds^\prime}{2 \pi \ii} \,\Big( \frac{\sqrt{\lambda}}{\pi} \Big)^{2s+2s^\prime+4} \,\Big( \frac{1}{2} \Big)^{2s+4s^\prime+4} \eta_{2s+1}\,\zeta_{2s^\prime+5}
    \notag \\
& \qquad\times ~\frac{\Gamma(k-s-1)\,\Gamma(k-s^\prime-1)\,\Gamma(2s+2)\,\Gamma(2s^\prime+6)}{\Gamma(s+k+2)\,\Gamma(s^\prime+k+2)}
\notag\\[2mm]
&=-\int_{-\ii \infty}^{+\ii \infty} \frac{ds}{2 \pi \ii}\int_{-\ii \infty}^{+\ii \infty} \frac{ds^\prime}{2 \pi \ii} \,\Big(\frac{\sqrt{\lambda}}{\pi} \Big)^{2s+2s^\prime+4}\,\Big( \frac{1}{2} \Big)^{2s+4s^\prime+4}\,\,\eta_{2s+1}\,\zeta_{2s^\prime+5}\notag \\ 
&\qquad \times~ \frac{\Gamma(-s)\,\Gamma(-s^\prime)\,\Gamma(2s+2)\,\Gamma(2s^\prime+6)}{2\,(s+s^\prime+2)\,\Gamma(s+2)\,\Gamma(s^\prime+2)}~.
\label{sumKYint2}
\end{align}
In closing the integration contours anti-clockwise we pick up two types of contributions: the one from the residues at $(s=-1\,,\,s^\prime=-1)$, and the one from the residues at $(s=-s^\prime-2\,,\,s^\prime=-n)$ for $n=1,2,\cdots$.
The first one is
\begin{align}
    -\frac{3}{2}\,\zeta_3~,
    \label{B1}
\end{align}
while the second contribution is formally
\begin{align}
    \frac{3}{2}\,\zeta_3-4\log 2+4\sum_{n=1}^\infty 4^{n}\,B_{2n}\,(4n^2-1)\,\eta_{2n+1}~.
    \label{B2}
\end{align}
Again the sum over $n$ is divergent but it can be regularized using the integral representation (\ref{zetaidentity}) of the Dedekind $\eta$-values and the generating function (\ref{Bernoulligen}) of the even Bernoulli numbers. In fact we have
\begin{align}
    4 \sum_{n=1}^\infty 4^{n}\,B_{2n}\,(4n^2-1)\,\eta_{2n+1}&= 
    4\int_0^\infty \frac{dt}{t}\,\frac{\rme^t\,t^2}{(\rme^t+1)^2}
    \sum_{n=1}^\infty
    \frac{(2n-1)\,B_{2n}}{(2n)!}\,(2t)^{2n}\notag\\
    &=
   4\int_0^\infty
    \frac{dt}{t}\,\frac{\rme^t\,t^2}{(\rme^t+1)^2} \,\Big[t\,\partial_t f(2t)-f(2t)\Big]
    =\frac{3}{2}\,\zeta_3~.
    \label{B3}
\end{align}
Using this in (\ref{B2}) and adding (\ref{B1}) we conclude that
\begin{align}
\cB\, \underset{\lambda \rightarrow \infty}{\sim} \, \frac{3}{2}\,\zeta_3 -4\log 2 ~.
\label{Bstrong}
\end{align}
Finally, combining (\ref{ZYappAB}), (\ref{Astrong}) and (\ref{Bstrong}), we simply get
\begin{align}
\sum_{k=1}^{\infty}(-1)^k\sqrt{2k}\,\mathsf{Z}_{2k}^{(4)}\,\mathsf{Y}_{2k} \,\underset{\lambda \rightarrow \infty}{\sim} \,\frac{3}{2}\,\zeta_3 ~.
\label{ZYstrongapp}
\end{align}
We have checked this result also numerically, finding agreement.

\section{The matrix model for the \texorpdfstring{Sp($N$)}{} theory}
\label{app:spn}

In this appendix, we briefly discuss the matrix model for the $\cN=2$ superconformal gauge theory with one anti-symmetric and four fundamental hypermultiplets. Even if this model has already been discussed
in the literature \cite{Beccaria:2021ism,Beccaria:2022kxy,Behan:2023fqq} we present here an analysis based on the use of the full Lie algebra approach with the purpose of showing that this methods also works quite efficiently in this case. We believe that this can help in better understanding and appreciating the similarities between the Sp($N$) theory and the \textbf{D} theory. In the sequel all quantities referring to the Sp($N$) theory will be denoted by a $~\widetilde{}~$ sign.

\subsection{The \texorpdfstring{$\widetilde{\mathcal{P}}$}{} operators}
By exploiting supersymmetric localization the partition function of the Sp($N$) theory can be written as an integral over a matrix $\widetilde{a}$ belonging to the Lie algebra $\mathfrak{sp}(N)$:
\begin{align}
\widetilde{\mathcal{Z}}_{\mathrm{Sp}} = \int d\widetilde{a} \,\,  \rme^{-\frac{8\p^2}{g^2}\tr
\widetilde{a}^2}\,|\widetilde{Z}_{\mathrm{1-loop}}\,\widetilde{Z}_{\mathrm{inst}}|^2~.    
\end{align}
In the large-$N$ 't Hooft limit we can neglect instanton corrections and set $\widetilde{Z}_{\mathrm{inst}}=1$.
The 1-loop part can be expressed as
\begin{align}
|\widetilde{Z}_{\mathrm{1-loop}}|^2 = \rme^{-\widetilde{S}_{\mathrm{Sp}}}\, ,
\end{align}
where
\begin{align}
\widetilde{S}_{\mathrm{Sp}} =  4\sum_{k=1}^{\infty} (-1)^{k+1} \Big(\frac{\lambda}{8\pi^2N}\Big)^{k+1}\,(2^{2k}-1)\,\frac{\zeta_{2k+1}}{k+1}\,\tr \widetilde{a}^{2k+2} \,  .
\label{SintSp}
\end{align}
We observe that in this matrix model there are no double-trace terms and that the above expression is formally identical to the single-trace action \eqref{Sintst} of the \textbf{D} theory with the replacement of $a$ by $\widetilde{a}$.

In analogy to what we have done in Section~\ref{sec:Dstar}, we introduce 
the $\mathcal{\widetilde{P}}$ operators defined as
\begin{align}
    \widetilde{\cP}_k=\sqrt{\frac{k}{2}}\,\sum_{\ell=0}^{\lfloor \frac{k-1}{2}\rfloor}(-1)^\ell\,\Big(\frac{N}{2}\Big)^{\ell-\frac{k}{2}}\,
    \frac{(k-\ell-1)!}{\ell!\,(k-2\ell)!}\,\Big(\tr \widetilde{a}^{k-2\ell}-
    \big\langle \tr \widetilde{a}^{k-2\ell}\big\rangle_0\,\Big)~.
    \label{Pntilde}
\end{align}
where $\langle ~ \rangle_0$ denotes the vacuum expectation value in the $\mathcal{N}=4$ SYM with gauge group $Sp(N)$. The definition \eqref{Pntilde} ensures that, at leading order in the large-$N$ expansion, the $\mathcal{\widetilde{P}}$ operators are orthonormal in the free theory:
\begin{align}
\big\langle \widetilde{\cP}_{2k_1}\,\widetilde{\cP}_{2k_2} \big\rangle_0 \, = \, \delta_{k_1,k_2} + O\Big(\frac{1}{N}\Big)~.   
\end{align}
Moreover, it can be shown that in the free theory the correlation functions among $\mathcal{\widetilde{P}}$ operators can be factorized à la Wick into product of 2- and 3-point correlators only.

Inverting \eqref{Pntilde}, we can express the traces of $\widetilde{a}$ in terms of the $\widetilde{\cP}$ operators and rewrite the interaction action \eqref{SintSp} as a sum of two contributions. Only the term which is linear in the $\widetilde{\cP}$ operators is relevant for our analysis, so that for all expectation values we can effectively use
\begin{align}
\widetilde{S}_{\mathrm{Sp}} =-\sum_{k=1}^{\infty}\widetilde{\textsf{Y}}_{2k}\,\mathcal{\widetilde{P}}_{2k}
\end{align}
where
\begin{align}
\widetilde{\mathsf{Y}}_{2k} = 4\,(-1)^{k+1} \,\sqrt{k}\int_0^\infty \!\frac{dt}{t}\,\frac{\rme^t}{(\rme^t+1)^2}\,J_{2k}\Big(\frac{\sqrt{\lambda}\,t}{\pi}\Big) -\delta_{k,1}\,\frac{\log 2}{2\pi^2}\,\lambda= \sqrt{2}\,\textsf{Y}_{2k}~.   
\end{align}
Following the same procedure discussed for the \textbf{D} theory, we compute the first orders of the large-$N$ expansion of the 1- and the 2-point connected correlators of the $\mathcal{\widetilde{P}}$ operators in the interacting Sp($N$) theory and get
\begin{align}
& \big\langle {\widetilde{\cP}}_{2n} \big\rangle_{\mathrm{Sp}} = \widetilde{\mathsf{Y}}_{2n} + \frac{\sqrt{n}}{2N}\,\widetilde{\mathsf{Y}}\,\Big(\widetilde{\mathsf{Y}}+\frac{1}{\sqrt{2}}\Big)+ O\Big(\frac{1}{N^2}\Big)~, \label{1ptSp} \\[1mm]
& \big\langle \widetilde{\cP}_{2n_1}\,\widetilde{\cP}_{2n_2} \big\rangle^{c}_{\mathrm{Sp}} = \delta_{n_1,n_2} + \frac{\sqrt{n_1n_2}}{2N}\,\Big(1+2\sqrt{2}\,\widetilde{\mathsf{Y}}\Big) + O\Big(\frac{1}{N^2}\Big)~, \label{2ptSp}
\end{align}
where
\begin{align}
\widetilde{\mathsf{Y}} = \sum_{k=1}^{\infty}\sqrt{2k}\,\,\widetilde{\mathsf{Y}}_{2k}~.    
\end{align}

\subsection{The mass-deformed theory}
Mimicking \cite{Behan:2023fqq} we consider a mass deformation of this Sp($N$) gauge theory, in which the four fundamental hypermultiplets gain a mass $\widetilde{m}_i$ ($i=1,\ldots,4)$.  We refer to this massive theory as Sp$^*$. Repeating the same steps of Section \ref{D*sec}, in the planar limit we can write the small-mass expansion of the partition function of the Sp$^*$ theory in the following form
\begin{align}
    \widetilde{\cZ}_{\mathrm{Sp}^*}=\int\!d\widetilde{a}\,\,\rme^{-\tr \widetilde{a}^2 }\,\rme^{-\widetilde{S}_{\mathrm{Sp}}-\sum_{i}\widetilde{m}_i^2\,\widetilde{S}_2-\sum_i \widetilde{m}_i^4\,\widetilde{S}_4+O(\widetilde{m}^6)}
    \label{ZSpstar}
\end{align}
where $\widetilde{S}_2$ and $\widetilde{S}_4$ are exactly defined as in \eqref{S2} and \eqref{S4} replacing $a\to\widetilde{a}$. Hence, for analogy with the \textbf{D}$^*$ theory, the quantities of our interest are
\begin{subequations}
    \begin{align}
    -\partial_{m_i}^4\cF_{\mathrm{Sp}^*}\Big|_{m=0}&=-24 \,\big\langle \widetilde{S}_4\big\rangle_\mathrm{Sp}+12\,
    \big\langle \widetilde{S}_2^{~2}\big\rangle_\mathrm{Sp}-12\,
    \big\langle \widetilde{S}_2\big\rangle_\mathrm{Sp}^2&& (i=1,\ldots,4)~,\label{d4fSp}\\[1mm]
   -\partial_{m_i}^2\partial_{m_j}^2\cF_{\mathrm{Sp}^*}\Big|_{m=0}&= 4\,
    \big\langle \widetilde{S}_2^{~2}\big\rangle_\mathrm{Sp}-4\,
    \big\langle \widetilde{S}_2\big\rangle_\mathrm{Sp}^2&&(i\neq j =1,\ldots,4)~.\label{d22fSp}
\end{align}
\label{Spderivatives}%
\end{subequations}
To evaluate the right-hand sides of these equations, the next step is to rewrite $\widetilde{S}_4$ and $\widetilde{S}_2$ in terms of the $\widetilde{\cP}$ operators introduced in \eqref{Pntilde}. Thus we have
\begin{align}
\widetilde{S}_4=\widetilde{S}_4^{(1)}+\widetilde{S}_4^{(0)}\qquad\mbox{and}\qquad \widetilde{S}_2=\widetilde{S}_2^{(1)}+\widetilde{S}_2^{(0)}
\label{S2S4tilde}
\end{align}
where, as usual, the superscripts $^{(1)}$ and $^{(0)}$ denote, respectively, the linear term in the $\widetilde{\cP}$ operators and the term coming from the vacuum expectation values $\langle \tr \widetilde{a}^{2k}\rangle_0$. Very similarly to what occurs in the \textbf{D}$^*$ theory, the right-hand sides of \eqref{S2S4tilde} 
can be expressed as
\begin{align}
\widetilde{S}_4^{(1)}&=-\frac{1}{6}\,\sum_{k=1}^\infty (-1)^k
\,\sqrt{k}\,\,\mathsf{Z}_{2k}^{(4)}\,\widetilde{\cP}_{2k}~,\label{S4ZSp}\\
\widetilde{S}_2^{(1)}&=2\sum_{k=1}^\infty (-1)^k\,\sqrt{k}\,\,\mathsf{Z}_{2k}^{(2)}\,\widetilde{\cP}_{2k}~, \label{S2ZSp}\\
\widetilde{S}_4^{(0)}&=-\frac{N}{3}\,\frac{2\pi}{\sqrt{\lambda}}\,\mathsf{Z}_1^{(3)}+\frac{1}{4}\zeta_3-\frac{\pi}{6\,\sqrt{\lambda}}\mathsf{Z}_1^{(3)}+\frac{1}{24}\mathsf{Z}_2^{(4)}-\frac{\lambda}{4608\,\pi^2 N}\,\,\mathsf{Z}_2^{(6)}+O\Big(\frac{1}{N^2}\Big)~,
    \label{S40ZSp}
\end{align}
where $\mathsf{Z}_n^{(p)}$ was defined in \eqref{Zkp}. 
We do not need to display the term $\widetilde{S}_2^{(0)}$ since in the connected combination $\langle \widetilde{S}_2^{~2}\rangle_\mathrm{Sp}-\langle \widetilde{S}_2 \rangle_\mathrm{Sp}^2$ appearing in (\ref{Spderivatives}) it cancels out.
To derive \eqref{S40ZSp} we used the fusion/fission identities of the Sp($N$) traces \cite{Huang2017GrouptheoreticRF} which lead to
\begin{align}
\big\langle \tr \widetilde{a}^{2k} \big\rangle_0  &= \frac{N^{k+1}}{2^{k-1}}\frac{(2k)!}{k!(k+1)!} + \frac{N^k}{2^k}\frac{(2k-1)!}{k!(k-1)!}(1-\delta_{k,0}) \notag \\
& \quad+ \frac{N^{k-1}}{2^{k+1}}\frac{(2k-1)(2k-3)!}{3(k-2)!(k-2)!}(1-\delta_{k,1})(1-\delta_{k,0}) + O(N^{k-2})~.
\label{tLarge}
\end{align}
Combining \eqref{S4ZSp} and (\ref{S40ZSp}) with the  1-point correlator \eqref{1ptSp} we obtain
\begin{align}
 \big\langle \widetilde{S}_4\big\rangle_\mathrm{Sp}&=-\frac{N}{3}\,\frac{2\pi}{\sqrt{\lambda}}\,\mathsf{Z}_1^{(3)} +\frac{1}{4}\zeta_3-\frac{\pi}{6\,\sqrt{\lambda}}\mathsf{Z}_1^{(3)}+\frac{1}{24}\mathsf{Z}_2^{(4)} -\frac{1}{6}\,\sum_{k=1}^\infty (-1)^k\,\sqrt{k}\,\,\mathsf{Z}_{2k}^{(4)}\,\widetilde{\mathsf{Y}}_{2k}\notag\\[1mm]
    &\qquad
    -\frac{1}{96\,N}\bigg[\frac{\lambda}{48\pi^2}\,\mathsf{Z}_2^{(6)}
    - \frac{\sqrt{\lambda}}{\pi}\,\mathsf{Z}_1^{(5)}\,\widetilde{\mathsf{Y}}\Big({\widetilde{\mathsf{Y}}}+\frac{1}{\sqrt{2}}\Big)\bigg]+O\Big(\frac{1}{N^2}\Big)~,
    \label{S4finSp}
\end{align}
while, exploiting \eqref{S2ZSp} and the connected 2-point correlator \eqref{2ptSp}, we find
\begin{align}
     \big\langle \widetilde{S}_2^{\,2}\big\rangle_\mathrm{Sp}-\big\langle \widetilde{S}_2\big\rangle_\mathrm{Sp}^2&=\sum_{k=1}^\infty 4k\,\big(\mathsf{Z}_{2k}^{(2)}\big)^2+
     \frac{1}{2\,N}\Big(\frac{\sqrt{\lambda}}{4\pi}\,\,\mathsf{Z}_1^{(3)}\Big)^2\,\left(1+2\sqrt{2}\,\widetilde{\textsf{Y}}\right)+O\Big(\frac{1}{N^2}\Big)~.
\label{S2finSp}
\end{align}
Comparing \eqref{S4finSp} and \eqref{S2finSp} with their counterparts in the \textbf{D} theory given in \eqref{S4fin} and \eqref{S2fin}, respectively, we see that the overall structure is similar but the coefficients are clearly different. Moreover, in the Sp($N$) gauge theory there is no contribution from the free energy of the \textbf{E} theory, since in the symplectic case the interaction action contains only the single-trace part. For this reason the perturbative expansions defined in \eqref{Spderivatives} are profoundly distinct from those given in \eqref{derivativesweak}. Indeed we have
\begin{subequations}
    \begin{align}
    -\partial_{m_i}^4\cF_{\mathrm{Sp}^*}\Big|_{m=0}~ &\underset{\lambda \rightarrow 0}{\sim}~N\big(24\,\zeta_3-120\,\zeta_5\,\hat\lambda+420\,\zeta_7\,\hat\lambda^2-1260\,\zeta_9\,\hat\lambda^3+\cdots\big)\label{Spd4Fweak}\\
    &\quad\quad-\Big[60\,\zeta_5\,\hat\lambda-
    (108\,\zeta_3^2+315\,\zeta_7)\,\hat\lambda^2-(720\,\zeta_3\,\zeta_5-1260\,\zeta_9)\,\hat\lambda^3+\cdots\Big]\notag\\[1mm]
    &\quad\quad+\frac{1}{N} \bigg[\Big(54\,\zeta_3^2+\frac{105}{2}\,\zeta_7\Big)\,\hat\lambda^2+(540\,\zeta_3\zeta_5-630\,\zeta_9)\,\hat\lambda^3+\cdots \bigg]+O\Big(\frac{1}{N^2}\Big)~,\notag\\[2mm]
   -\partial_{m_i}^2\partial_{m_j}^2\cF_{\mathrm{Sp}^*}\Big|_{m=0}~ &\underset{\lambda \rightarrow 0}{\sim}~\big(36\,\zeta_3^2\,\hat\lambda^2-240\,\zeta_3\,\zeta_5\,\hat\lambda^3+\cdots\big)\notag\\
    &\quad\quad+\frac{1}{N} \big(18\,\zeta_3^2\,\hat\lambda^2-180\,\zeta_3\zeta_5\,\hat\lambda^3+\cdots\big)+O\Big(\frac{1}{N^2}\Big) ~, \label{Spd22Fweak}
\end{align}
\label{Spderivativesweak}%
\end{subequations}
where $\hat\lambda=\lambda/(8\pi^2)$.

Quite remarkably, at strong coupling the same quantities behave very similarly. This is due to the strong-coupling expansion of the $\mathsf{Z}_k^{(p)}$ coefficients and of $\widetilde{\mathsf{Y}}$. Exploiting the results derived in Appendix \ref{app:strong}, it is straightforward to check that 
\begin{subequations}
\begin{align}
-\partial_{m_i}^4\cF_{\mathrm{Sp}^*}\Big|_{m=0} ~  & \underset{\lambda \rightarrow \infty}{\sim} ~ \frac{32\pi^2}{\lambda}N +6\log\lambda +12\gamma - 12\log(4\pi) -12\,\zeta_3 +22
\notag \\[1mm] & 
\qquad\qquad+ \frac{3}{N}\Big(1-\frac{\log 2 }{\pi^2}\,\lambda \Big)+
O\Big(\frac{1}{N^2}\Big)~, \label{Spd4Fstrong}\\[2mm]
-\partial_{m_i}^2\partial_{m_j}^2\cF_{\mathrm{Sp}^*}\Big|_{m=0}~& \underset{\lambda \rightarrow \infty}{\sim} ~ 2\log\lambda +4\gamma -4\log (4\pi)-4\,\zeta_3+\frac{22}{3} \nonumber\\
&\qquad\qquad +\frac{1}{N}\Big(1-\frac{\log 2}{\pi^2}\,\lambda\Big) +O\Big(\frac{1}{N^2}\Big) ~.
\label{Spd22Fstrong}
\end{align}
\end{subequations}%
Up to order $1/N$, our results agree with the expressions in Eq.\,(3.13) of \cite{Behan:2023fqq}.

\providecommand{\href}[2]{#2}\begingroup\raggedright\endgroup

\end{document}